\documentclass[reprint, pra,showpacs,amsmath,amssymb,superscriptaddress,showkeys,longbibliography]{revtex4-1}

\usepackage{hyperref}
\usepackage{braket}
\usepackage{graphicx}
\usepackage{xcolor}

\begin{document}

\title{Quantum-fluctuation asymmetry in multiphoton Jaynes-Cummings resonances}

\author{Th. K. Mavrogordatos}
\email[Email address: ]{themis.mavrogordatos@fysik.su.se}
\affiliation{Department of Physics, Stockholm University, SE-106 91, Stockholm, Sweden}
\affiliation{ICFO -- Institut de Ci\`{e}ncies Fot\`{o}niques, The Barcelona Institute of Science and Technology, 08860 Castelldefels (Barcelona), Spain} 

\date{\today}

\begin{abstract}
We explore the statistical behaviour of the light emanating from a coherently driven Jaynes-Cummings (JC) oscillator operating in the regime of multiphoton blockade with two monitored output channels causing the loss of coherence at equal rates. We do so by adopting an operational approach which draws the particle and wave aspects of the forwards scattered radiation together, building upon the relationship between quantum optical correlation functions and conditional measurements. We first derive an analytical expression of the intensity cross-correlation function at the peak of the two-photon JC resonance to demonstrate the breakdown of detailed balance. The application of quantum trajectory theory in parallel with the quantum regression formula subsequently uncovers various aspects of temporal asymmetry in the quantum fluctuations characterizing the cascaded process through which a multiphoton resonance is established and read out. We find that monitoring different quadratures of the cavity field in conditional homodyne detection affects the times waited between successive photon counter ``clicks'', which in turn trigger the sampling of the homodyne current. Despite the fact that the steady-state cavity occupation is of the order of a photon, monitoring of the developing bimodality also impacts on the ratio between the emissions directed along the two decoherence channels.  
\end{abstract}


\maketitle

\section{Introduction}
 
Multiphoton processes express the quantum mechanical substitute of a classical nonlinearity accompanied by the presence of large fluctuations in an open system which prevent the reduction of its dynamical evolution to a simple set of transport-like equations featuring in classical or at best a semiclassical treatment~\cite{CarmichaelQO1, CarmichaelQO2, Tian1992}. The strong-coupling ``thermodynamic limit'' related to the persistence of photon blockade in the open coherently driven Jaynes-Cummings (JC) model exemplifies in a very characteristic way the impossibility of reducing quantum dynamics into a description of ``fuzz'' on top of a mean-field solution. Energy-level shifts due to one, two, three and more individual photons are fundamental to photon blockade~\cite{Birnbaum2005, Hamsen2017, Najer2019}; not only are the multiphoton resonances totally missed by the mean-field nonlinearity, but the disagreement between the latter and the full quantum treatment via the source master equation (ME) does not disappear as the system size parameter is sent to infinity~\cite{PhotonBlockade2015}. While the possibility of observing photon antibunching in an $m$-photon process has been highlighted since 1980~\cite{Zubairy1980} and is typically associated with photon blockade, Shamailov and collaborators noted that both bunching (possibly extreme) and antibunching may be observed in the readout of a two-photon resonance, depending on the drive amplitude~\cite{Shamailov2010}; see also the cooperative resonance fluorescence from two atoms~\cite{Richter1982}.  

In an explicitly open-system configuration, when characterizing the emission of $N$-photon bundles in the driven JC model situated far in the dispersive regime and for high excitation~\cite{Munoz2014}, S\'anchez Mu\~noz and coworkers observe that ``for time windows larger than the coherence time, counting of the photon bundles becomes Poisson distributed, as short-time correlations are lost''. In this example, the description of light-matter interaction can be reduced to the four-level ladder pertaining to the Mollow triplet we encounter in resonance fluorescence: the laser field dresses the two-state `atom' while the presence of the cavity generated a Purcell enhancement of the transition rate between two different atomic manifolds. To bring in the wave aspect of radiation, time-asymmetric amplitude-intensity correlations were predicted for the incoherently scattered light by certain transitions of single three-level and $V$-shaped atoms~\cite{Xu2015}, generalizing the experimental results that had been obtained a few years earlier of interference measurements conditioned on the detection of a fluorescence photon~\cite{Gerber2009}. While resonance fluorescence features non-Gaussian fluctuations, detailed balance is imposed by the low dimensionality, since the resonant scattering involves transitions between only two states~\cite{Klein1955, Denisov2002}. Moreover, the distinct role of quantum interference in $V$-type media was elucidated via a quantum-trajectory analysis in~\cite{Carmichael1997}.   

Detailed balance is not guaranteed to hold away from equilibirum~\cite{Tomita1973, Tomita1974, Klein1955}. Despite the fact that the laser is a characteristic system out of equilibrium which satisfies detailed balance~\cite{Graham1971}, time-asymmetric correlations have been reported in several configurations of cavity QED since the early 2000s~\cite{GiantViolations2000, Foster2000, Denisov2002, CarmichaelFosterChapter}. For a cavity-QED source, where $a^{\dagger}$ and $a$ are creation and annihilation operators for the field, the time- and normal-ordered correlation function
\begin{equation}\label{eq:Hdef}
 2 H_{\theta}(\tau)= \begin{cases}
                      e^{-i\theta}\langle a^{\dagger}(t)a(t+\tau)a(t)  \rangle + \text{c.c.},\quad \tau\geq 0\\
                      e^{-i\theta}\langle a^{\dagger}(t-\tau)a(t-\tau)a(t) \rangle + \text{c.c.},\quad \tau\leq 0,
                     \end{cases}
\end{equation}
is measured in a single channel of intensity and field amplitude through conditional homodyne detection with a local oscillator of phase $\theta$~\cite{GiantViolations2000, CarmichaelFosterChapter}. When one is dealing with Gaussian fluctuations, $H_{\theta}(\tau)$ reduces to the autocorrelation function of the field amplitude, which is time-symmetric by definition. Consequently, a time-asymmetric correlation not only ascertains the breakdown of detailed balance, but also provides a direct indication of non-Gaussian fluctuations~\cite{Denisov2002}.  

The role of spontaneous emission in degrading the squeezing signal obtained from a resonantly excited cavity QED source was addressed  with the help of quantum trajectories in Ref.~\cite{Reiner2001}. Here the JC oscillator will be excited off-resonance in its strong-coupling limit. We will find that the various quantum jumps down the two decoherence channels do not appear on an equal footing when we monitor a multiphoton resonance in the course of a single realization. In direct photodetection, their difference is assessed against their ability to induce a conditional separation of the timescales predicted by the equations of motion for the unconditional dynamics: being most obvious in the two-photon resonance, individual quantum jumps revive or suppress the quantum beat against the semiclassical ringing associated with the saturation of the two-photon transition~\cite{Mavrogordatos2024}. The timescales themselves, coexisting in the solution of the ME, represent a merge of the classical nonlinearity attributed to the saturation of a two-level transition and the inherently quantum features of the cascaded process involving the JC spectrum [see Figs. 3 (a,b) of~\cite{Bishop2009} for an exemplary emergence] and being revealed in the strong-coupling limit of QED. Among the latter and for the two-photon resonance, one finds a characteristic quantum interference between the intermediate states of a four-state ladder, taking the form of a fast oscillation (beat) at twice the light-matter coupling strength~\cite{Shamailov2010, Lledo2021}. 

In this report, we aim to uncover the conditional dynamical evolution of photon-emission probabilities alongside the associated field and particle statistical properties operationally determined in the formation and readout of a JC multiphoton resonance for the ``time windows larger than the coherence time'' of~\cite{Munoz2014}. The multiple facets of quantum-fluctuation asymmetry play a key role in this journey and are intertwined with the dual nature of the monitored light. In Sec.~\ref{sec:particlecorsscorr}, we are dealing with a time asymmetry pertaining to the particle aspect of light monitored by the two distinct output channels with impedance matching conditions. We discuss the ``vacuum'' Rabi against the two-photon resonance, to mark the passage from the spectrum of a linearly-coupled oscillator formulation to the nonlinear $\sqrt{n}$ nonlinearity underlying incoherent scattering. Next, in Sec.~\ref{sec:WPcorr}, we employ the wave-particle correlator and modify the monitoring channels of the JC output in order to address the question whether the different quadratures of the cavity field (wave aspect) impact on the ratio of spontaneous vs. cavity emissions and on the times waited between the cavity emissions that trigger the homodyne-current sampling (particle aspect). As we follow along our way to give an affirmative answer, we meet up with an anomalous phase switching of the conditioned cavity-field amplitude, while the conditioned Wigner functions of the cavity field record a quantum interference of dressed JC states which is dynamically modified in the course of the cascaded process underlying a multiphoton resonance. Short concluding observations on the various aspects of quantum-fluctuation asymmetry close the paper out.   

\section{Intensity cross-correlation at the ``vacuum'' Rabi vs. two-photon resonance}
\label{sec:particlecorsscorr}

A ``vacuum'' Rabi splitting~\cite{CarmichaelQO2, Zhu1990} is the fundamental origin of the oscillations occurring in the intensity correlation function extracted from the output of two linear coupled and damped harmonic oscillators; although the correlation can be calculated from the ME and the quantum regression formula~\cite{Carmichael1986}, the splitting essentially retains a classical origin~\cite{CarmichaelClassical1991}. In this section, we will compare the ``vacuum'' Rabi resonance against the two-photon resonance to operationally deduce how does the paradigmatic JC nonlinearity differ from the model of two linear coupled oscillators. We begin our discussion on the breakdown of detailed balance by considering the cross-correlation of side and forwards emissions at the peak of the two-photon resonance. 

The JC Hamiltonian modeling the coherently-driven source in the unraveling schemes we will be dealing with takes the familiar form (in the interaction picture) 
\begin{equation}
H_{\rm JC}=-\hbar\Delta\omega_d(\sigma_{+}\sigma_{-} + a^{\dagger}a)+\hbar g(a\sigma_{+}+a^{\dagger}\sigma_{-}) + \hbar \varepsilon_d(a+a^{\dagger}), 
\end{equation}
where $\sigma_{+}$ and $\sigma_{-}$ are raising and lowering operators for the two-state atom, $g$ is the dipole coupling strength, $\varepsilon_d$ the drive amplitude, and $\Delta\omega_d$ the drive detuning. The reduced system density matrix $\rho$ obeys a standard Lindblad ME~\cite{PhotonBlockade2015, CarmichaelQO2} with two decoherence channels; the damping rates of the cavity field and the atomic polarization are $\kappa$ and $\gamma/2$, respectively. The operating regime of the open driven JC model we consider is defined by the following hierarchy of scales:
\begin{equation}
\varepsilon_d/g \ll 1, \quad \Delta\omega_d \sim g, \quad \gamma/g \ll 1,
\end{equation}
while we take $\gamma=2\kappa$ throughout--an impedance matching condition between the two channels~\cite{Carmichael1986}. The quantum-trajectory analysis of Ref.~\cite{Tian1992}, carried out as well for impedance matching conditions, has identified an asymmetry in the Mollow-triplet spectrum obtained down the two separate channels when studying the dynamic Stark splitting at the ``vacuum'' Rabi resonance. The asymmetry is a direct consequence of the transitions involving the unequally spaced multilevel JC spectrum even at low excitation of the targeted ``two-level'' system.  

Multiphoton resonances involve the JC dressed states in the manifold comprising the ground state $|\xi_0\rangle=|0, -\rangle$ and the excited couplet states $|\xi_{n}\rangle=\tfrac{1}{\sqrt{2}}(|n, -\rangle-|n-1,+\rangle)$ and $|\xi_{n+1}\rangle=\tfrac{1}{\sqrt{2}}(|n, -\rangle+|n-1,+\rangle)$ for $n\geq 1$, where $|n,\pm\rangle \equiv |n\rangle |\pm \rangle$, with $|\pm\rangle$ the upper and lower states of the two-state atom, and $\ket{n}$ the Fock states of the cavity field. In this section we operate at the peak of the two-photon JC resonance with $\Delta\omega_d/g=-1/\sqrt{2} + \mathcal{O}((\varepsilon_d/g)^2)$ where the effective two-photon coupling strength is $\Omega/g=2\sqrt{2}(\varepsilon_d/g)^2$ to dominant order. The resonance is established between $|\xi_0 \rangle$ and $|\xi_3 \rangle$, while the states $|\xi_1\rangle, |\xi_2 \rangle$ play a crucial part in the cascaded decay~\cite{Shamailov2010, Lledo2021}. 

The intensity cross correlation between the photons lost from the cavity (subsystem and channel A: resonant cavity mode and cavity decay) and the photons lost to the modes of the vacuum other than the preferential cavity mode (subsystem and channel B: two-state atom and spontaneous emission), derived in Appendix~\ref{sec:4levelcross}, takes the following time-asymmetric form: 
\begin{widetext}
\begin{equation}\label{eq:crosscorr}
g^{(2)}_{AB}(\tau)=\begin{cases}1+\tfrac{1}{15}e^{-2\gamma \tau} -\tfrac{7}{15}\tfrac{\gamma}{\Omega}e^{-\gamma \tau}\sin(2\Omega\tau) + \tfrac{1}{15}\left(3\tfrac{\gamma^2}{\Omega^2} -4\right)e^{-\gamma \tau}\cos(2\Omega\tau)-\tfrac{1}{15}\tfrac{\gamma^2+4\Omega^2}{\Omega^2}e^{-\gamma\tau}\cos(\nu\tau), \quad  \tau\geq 0\\
 1+\tfrac{1}{15}e^{-2\gamma |\tau|} -\tfrac{7}{15}\tfrac{\gamma}{\Omega}e^{-\gamma |\tau|}\sin(2\Omega|\tau|) + \tfrac{1}{15}\left(\tfrac{\gamma^2}{\Omega^2} -12\right)e^{-\gamma  |\tau|}\cos(2\Omega\tau)+\tfrac{1}{15}\tfrac{\gamma^2+4\Omega^2}{\Omega^2}e^{-\gamma|\tau|}\cos(\nu\tau), \quad  \tau\leq 0.
 \end{cases}
\end{equation}
\end{widetext}
The above expression is to be contrasted with the temporal symmetry of an intensity (auto)correlation function, which for the peak of the two-photon resonance assumes the form of Eq. 44 in~\cite{Shamailov2010}. 
\begin{figure*}
\centering
 \includegraphics[width=\textwidth]{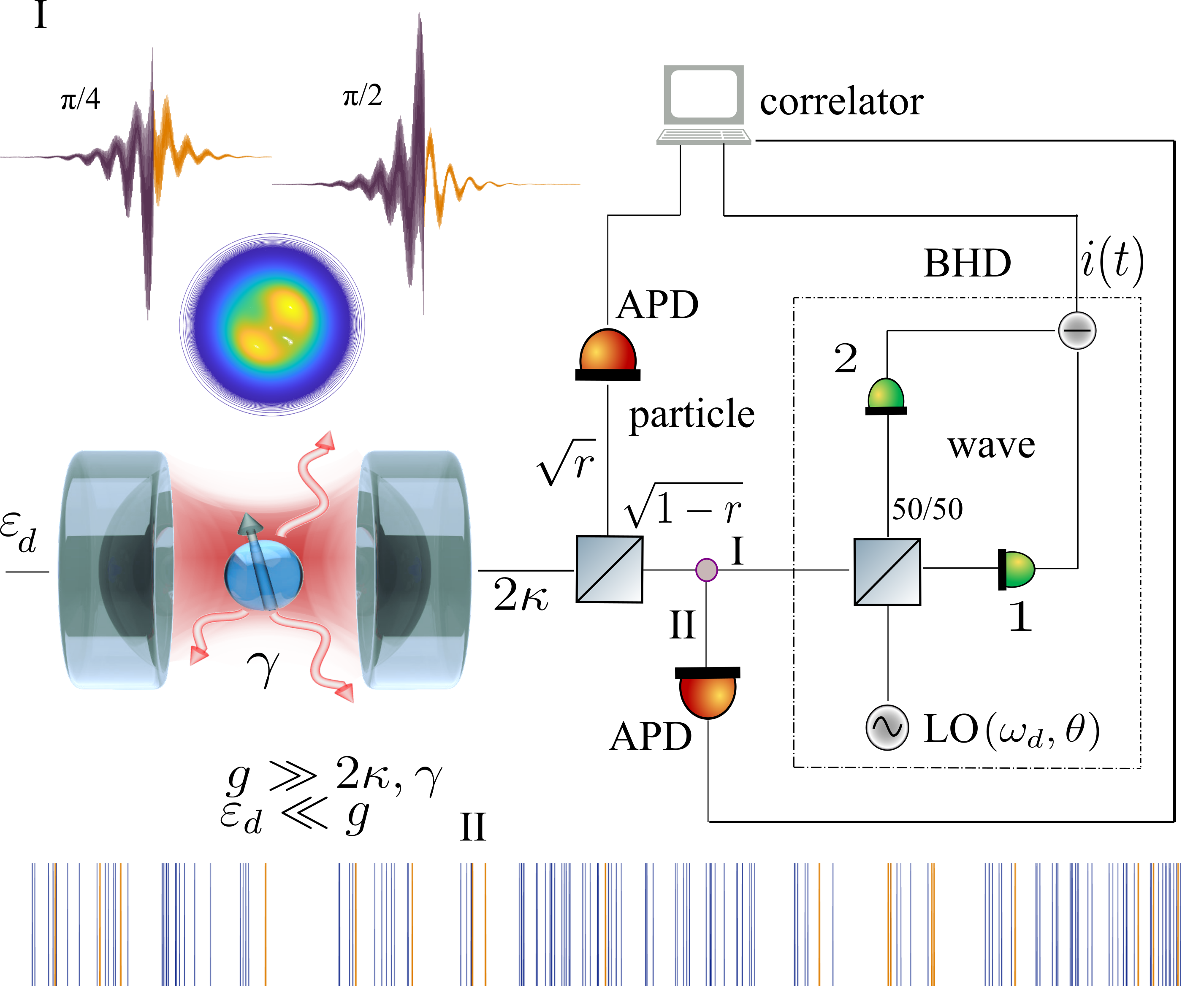}
 \caption{{\it Schematic representation of the Hanbury-Brown Twiss and the wave-particle correlators} for the JC (adapted from Fig. 1 of~\cite{Settineri2021}) photoemissive source. The choice between the two paths I (II) is made by, {\it e.g.}, the absence (presence) of a perfect mirror at the point marked by the thick dot after the first beam splitter. Path I is taken to generate a wave-particle correlation $H_{\theta}(\tau)$, shown in the time-asymmetric plots (the two colors distinguish negative from positive time delays) at the top left of the figure for a two-photon JC resonance unraveling with $\theta=\pi/4$ and $\pi/2$. The contour plot underneath depicts the Wigner function of the steady-state bimodal cavity field. Along this path, a fraction $1-r$ of the input photon flux is guided towards a balanced homodyne detector (BHD) generating a photocurrent $i(t)$, while the rest (fraction $r$) is sent to a photon counter [an avalanche photodiode (APD)] triggering the sampling of $i(t)$. The local oscillator (LO) has a variable $\theta$, while its frequency is locked to $\omega_d$. The driving amplitude is $\varepsilon_d/g=0.5$, and the detuning is calculated from the perturbative formula $\Delta\omega_d/g=1/\sqrt{2}+\sqrt{2}(\varepsilon_d/g)^2$. Path II is taken with $r=1/2$ to generate the intensity correlation function $g^{(2)}(\tau)$ in a typical Hanbury-Brown and Twiss arrangement. A sample photoelectron record is shown at the bottom for a direct-photodetection unraveling of the seven-photon JC resonance over tens of inverse cavity lifetimes [which we meet again in Fig.~\ref{fig:fig5}(a)].}
 \label{fig:fig1}
\end{figure*}

The intensity cross-correlation of the output photons collected when driving at the ``vacuum'' Rabi resonance appears to inherit the symmetry of $g^{(2)}(\tau)$, as shown in Fig.~\ref{fig:figAppCorr} of App.~\ref{sec:4levelcross}. Slight deviations from a fully symmetrical profile are observed owing to the occupation of couplet states higher than $|\xi_0 \rangle$ and $|\xi_1\rangle$ -- with unequal matrix elements for $a$ and $\sigma_{-}$ -- giving rise to a weak quantum beat on top of the semiclassical ringing (see also the waiting-time distributions of Ref.~\cite{Tian1992}). It bears emphasis that the period of the cross-correlations is also time-asymmetric. Driving at the two-photon resonance with the same strength amplifies the quantum beat (superimposed now on top of a lower-frequency oscillation), and brings about a pronounced temporal asymmetry in the cross correlation. 

Let us briefly focus on resonant excitation ($\Delta\omega_d=0$) in the weak-driving field limit $\varepsilon_d/g \sim \kappa/g$. We take $g^2/(\kappa\gamma)\gg 1$ in Eq. (9) of~\cite{Denisov2002} [in most cases considered here we operate with $g/\kappa \lesssim 10^3$] to obtain:
\begin{widetext}
\begin{equation}
 g^{(2)}_{AB,{\rm res}}(\tau)=\left\{ 1 + e^{-\gamma|\tau|/2}\left[\cos(g\tau)-(2g/\gamma-(\gamma/2g){\rm sgn}(\tau))\sin(g\tau) \right]\right\}^2 \approx \left[1 - e^{-\gamma|\tau|/2}(2g/\gamma)\sin(g\tau) \right]^2,
\end{equation}
\end{widetext}
which shows that the time asymmetry is inherited from an odd high-amplitude $\propto \sin(g \tau)$ dependence on top of the steady-state intensity correlation. This term is responsible for a highly oscillatory profile with an exponentially decaying envelope peaking at $g^{(2)}_{AB,r}(0)\approx (2g/\gamma)^2$. The zeros of the Rabi oscillation are located at $\gamma \tau_m \approx m\pi \gamma/g$, $m=\ldots -2,-1,0,1,2,\ldots$, in contrast to the asymmetric cross-correlation for the JC multiphoton resonances which never falls to zero for positive delays. Instead, this latter is bounded from below by the exponential decay $g^{(2)}_{AB}(0)e^{-\gamma \tau}$. In fact, for $\varepsilon_d/g \sim \gamma/g$ and $g/\kappa \sim 10^3$, the zero-delay cross-correlation may reach values as high as $g^{(2)}_{AB}(0)\sim 10^2$, in line with the extreme photon bunching noted in Ref.~\cite{Shamailov2010}. We are here dealing with the regime of amplitude quadrature squeezing in a direction set by the drive (here $\theta=\pi/4$), observed before the onset of bimodality. Since the azimuthal symmetry of the phase-space distribution is broken in the build-up to the steady state, but is restored after a photon emission event down either of the two channels, the transient evolution solving the ME will involve a quantum interference of dressed JC states with variable weights~\cite{Mavrogordatos_2023}. We aim to unravel such a quantum interference into single quantum trajectories by selecting different environments that might be encountered by the scattered field, all consistent with the validity of the ME~\cite{Carmichael1999}. For that, we will analyze trajectories that go into making up a second-order correlation, such as $g^{(2)}_{AB}$ of Eq.~\eqref{eq:crosscorr} or $g^{(2)}(\tau)$, as well as a third-order correlation, as is $H_{\theta}(\tau)$ of Eq.~\eqref{eq:Hdef}. In the next section, we will see how can these scattering records be obtained and what information do they provide. 

\section{The wave-particle correlator: exploring complementary unravelings}
\label{sec:WPcorr}

We now move to examine a cross-correlation of different observables along a {\it single} output channel, along the lines of Eq.~\eqref{eq:Hdef}, uncovering the tensions raised by the wave/particle duality of light~\cite{CarmichaelTalk2000}. As we will find later on, for the two-photon resonance operation and with no coherent offset added to the cavity output field~\cite{GiantViolations2000, CarmichaelFosterChapter}, deviations from the minimal four-state model of Sec.~\ref{sec:particlecorsscorr} and App.~\ref{sec:4levelcross}-- predicting a zero average field whence a vanishing $H_{\theta}(\tau)$ -- give rise to a nonzero wave-particle correlation, which shows pronounced oscillations at a timescale lying between the quantum beat and the semiclassical ringing. The difference from a vanishing correlation might have been expected by a symmetry breaking with respect to a diagonal (as set by the drive, here $\theta=\pi/4$) for the Wigner distribution of the cavity field~\cite{Wigner2PB} in its standard definition~\cite{CarmichaelQO1}
\begin{widetext}
\begin{equation}
 W(x+iy,x-iy;t)=\frac{1}{\pi^2}\int_{-\infty}^{\infty}d\mu \int_{-\infty}^{\infty}d\nu \chi_S(\mu+i\nu, \mu-i\nu;t)e^{-2i(\mu x-\nu y)},
\end{equation}
\end{widetext}
where $\chi_S(\mu+i\nu,\mu-i\nu)\equiv {\rm tr}\{\rho_{\rm cav}(t)e^{i[(\mu+i\nu)a^{\dagger}+(\mu-i\nu)a]}\}$ is the symmetrically ordered characteristic function, and $\rho_{\rm cav}$ is the cavity density matrix obtained after tracing out the `atomic' degrees of freedom.

Correlations are operationally determined via the {\it wave-particle} (alternatively {\it field-amplitude---intensity}) correlator through the setup depicted in Fig.~\ref{fig:fig1}. This device was first proposed in~\cite{GiantViolations2000} as an extension of the intensity cross-correlation of light introduced by Hanbury-Brown and Twiss~\cite{Brown1956, BrownHTwissI,BrownHTwissII}, and was shortly after used to record the conditional time evolution of a cavity field of a fraction of a photon in the experiment of~\cite{Foster2000}. A connection between the wave-particle correlation and weak measurements was pointed out in~\cite{Wiseman2002}. Fluctuations of the output light field emanating from the cavity are measured with a balanced homodyne detector (BHD), while the remaining fraction $r$ of the output photon flux is directed towards a photon counter which triggers the sampling of the homodyne current. Between starts, the conditional wavefunction evolves continuously under a stochastic Schr\"{o}dinger equation with a non-Hermitian Hamiltonian. This evolution is conditioned on the ongoing realization of the charge deposited in the homodyne detector output circuit. 

To begin with a first demonstration of the breakdown of detailed balance that can be inferred from the setup of Fig.~\ref{fig:fig1}, in the Appendices~\ref{sec:4levelcross} and~\ref{sec:thewp} we extract the intensity cross-correlation $g_{AB}^{(2)}(\tau)$ and the wave-particle correlation $H_{\theta}(\tau)$ for the scattering occurring at the ``vacuum'' Rabi resonance ($|\Delta\omega_d/g|=1$). We find that $H_{\theta}(\tau)$ displays a pronounced temporal asymmetry of fluctuations when the transition saturates as $\theta \to 0$. The correlations also vary in amplitude when other quadratures are probed. Furthermore, the selection of a different quadrature modulates a high-frequency oscillation--the quantum beat we met in Sec.~\ref{sec:particlecorsscorr}. 

\subsection{Scanning through the Jaynes-Cummings multiphoton resonances}

In Fig.~\ref{fig:fig2}, we unravel the buildup of multiphoton resonances into single realizations evincing the change in photon counting records with increasing cavity excitation. The top left frame of Panel I illustrates the development of multiphoton resonances through the steady-state solution of the ME with variable detuning. Reproducing the familiar pattern from Refs.~\cite{Shamailov2010, PhotonBlockade2015}, the low-order resonances saturate and broaden with increasing excitation, while their peaks are displaced in a nonlinear fashion when varying the drive amplitude, owing to the so-called {\it dressing of the (JC) dressed states}~\cite{Tian1992, CarmichaelQO2}. The frame underneath presents a single realization for the conditioned intracavity photon number $\langle a^{\dagger}a(t) \rangle_{\rm REC}$ obtained from a scan of the detuning to access progressively higher-order multiphoton resonances for the highest drive amplitude selected in the upper frame (green line). With the sole exception of the two-photon resonance peak, the photon emission in the forwards direction is bunched in the selected detuning range, while the zero-delay intensity (auto-)correlation $g^{(2)}(0)$ shows dips at the multiphoton resonance peaks. We have selected the complementary unraveling scheme of heterodyne detection [see Appendix~\ref{sec:hetdet}] to illustrate the multi-stability associated with the three- and seven- photon resonance peaks. A superposition of quantum beats is constantly present on top of the switching between multiple meta-stable states. Especially at the seven-photon resonance peak, from the single realization depicted we observe intense quantum-fluctuation switching among metastable states with different photon occupation, to produce an average over the rather modest steady-state photon number $\langle a^{\dagger}a \rangle_{\rm ss}\approx 1.83$; we will see later on that in the complementary unraveling of direct photodetection [Fig.~\ref{fig:fig5}(a)], these peaks to the higher rungs of the ladder will serve as signs of photon bunching. 
\begin{figure*}
\centering
 \includegraphics[width=\textwidth]{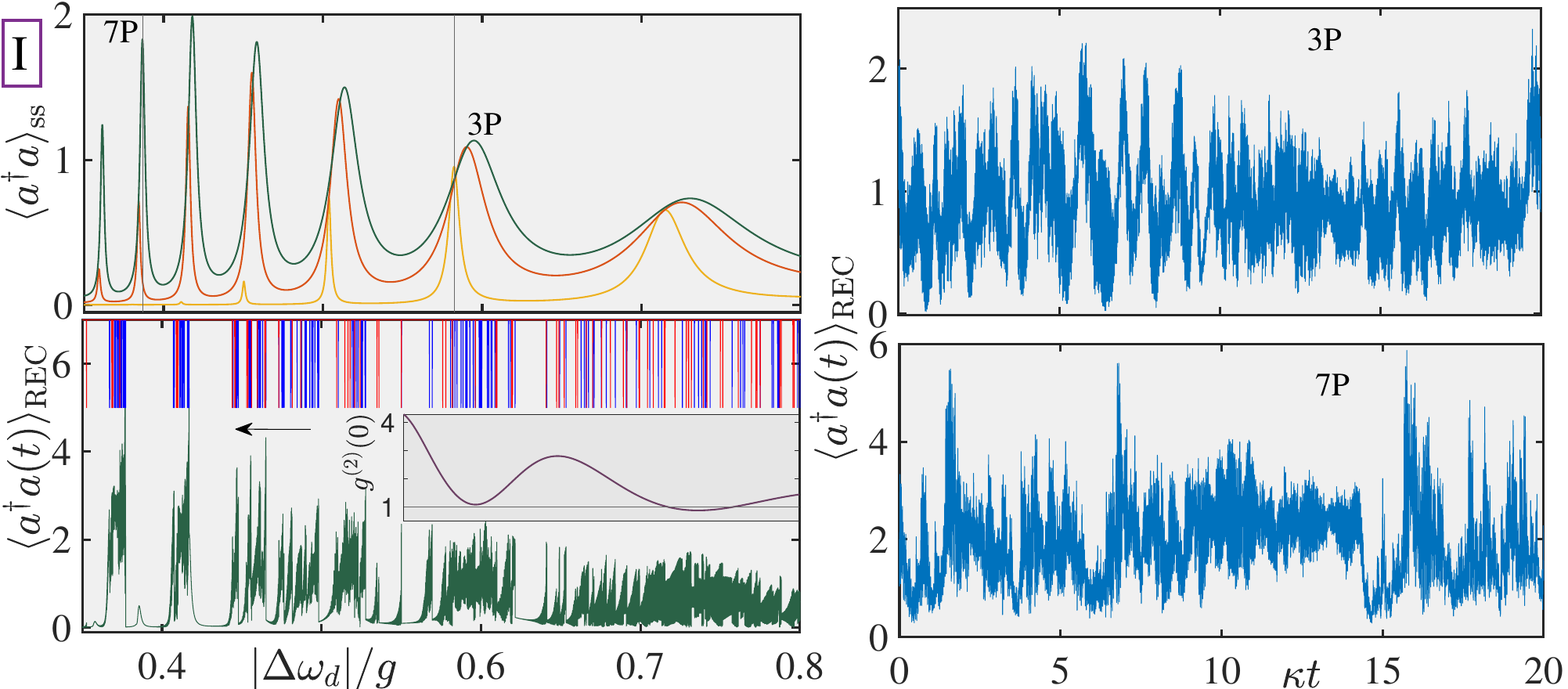}
  \includegraphics[width=\textwidth]{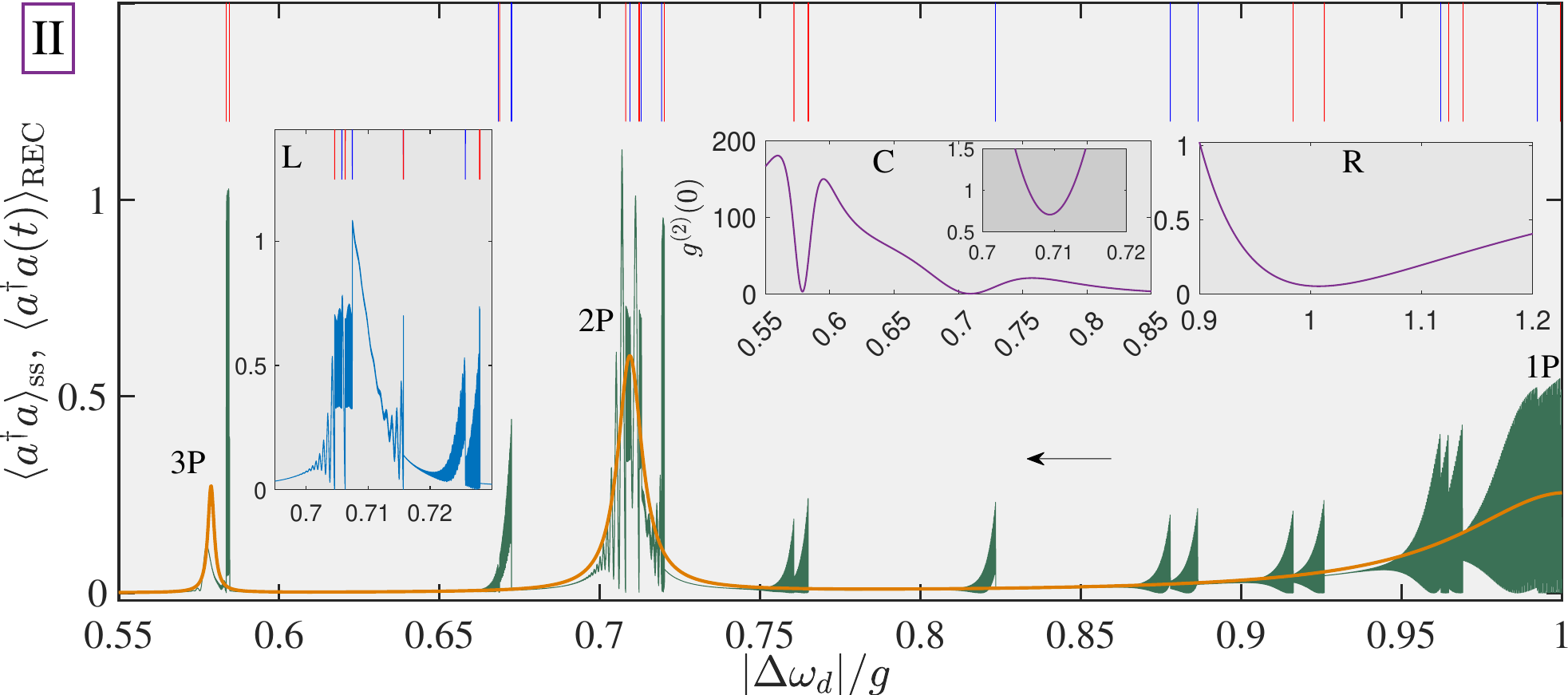}
   \includegraphics[width=\textwidth]{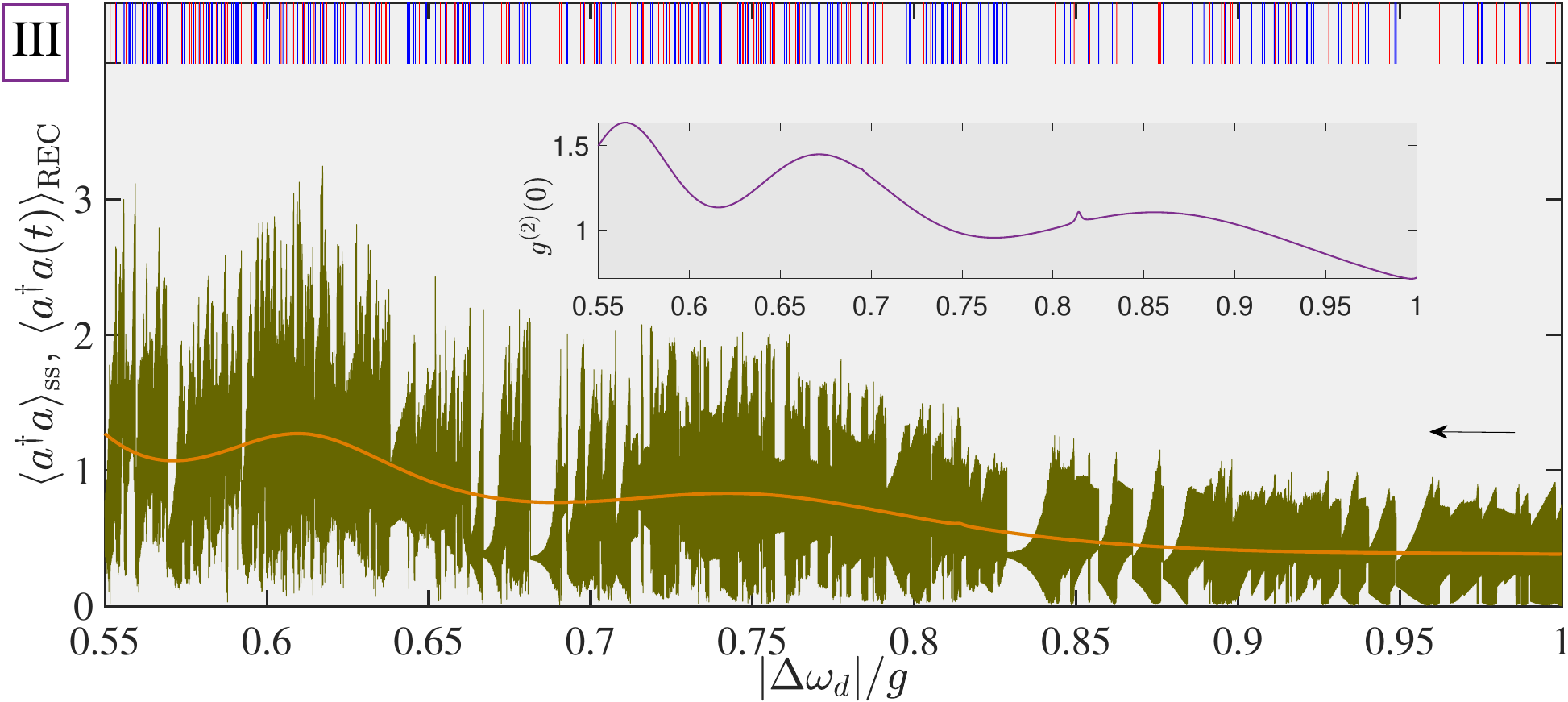}
\end{figure*}
\begin{figure*}
    \caption{{\it Visualizing the collapse of photon blockade in single realizations.} \underline{Panel I:} (Top left) Steady-state cavity photon number $\langle a^{\dagger}a \rangle_{\rm ss}$ for three different values of the drive amplitude $\varepsilon_d/g$: $0.075, 0.12, 0.14$ for increasing excitation (orange $\to$ ref $\to$ green). (Bottom left) Fluctuating conditional photon number $\langle a^{\dagger}a(t) \rangle_{\rm REC}$ for a scan of the detuning $\Delta\omega_d/g$, traversing the interval $[0.35, 1]$ (the frame zooms on a subset) in the duration of $\kappa T=200$, along the direction indicated by the black arrow. The drive amplitude is set to $\varepsilon_d/g=0.14$. The framed inset depicts the steady-state intensity correlation function at zero delay $g^{(2)}(0)$ for a detuning varying in the interval $[0.55, 0.8]$. Strokes in red (blue) indicate spontaneous (cavity) emissions. The two frames on the right depict the fluctuating conditional photon number under heterodyne detection, at the peaks of the three-(3P) and seven-(7P) photon resonances, indicated at the top left frame. \underline{Panels II, III:} Same as for the bottom left frame of I, but for $\varepsilon_d/g=0.04$ and $0.2$, respectively [the framed insets C(enter) and R(right) in II plots $g^{(2)}(0)$]. The (L)eft inset in II focuses on the two-photon resonance peak in a different realization with a finer time discretization. The solid orange line depicts the steady-state photon number obtained from the numerical solution of the ME. The labels 2P and 3P mark the two- and three-photon resonance peaks, respectively.}
     \label{fig:fig2}
\end{figure*}

\begin{figure*}
\centering
 \includegraphics[width=\textwidth]{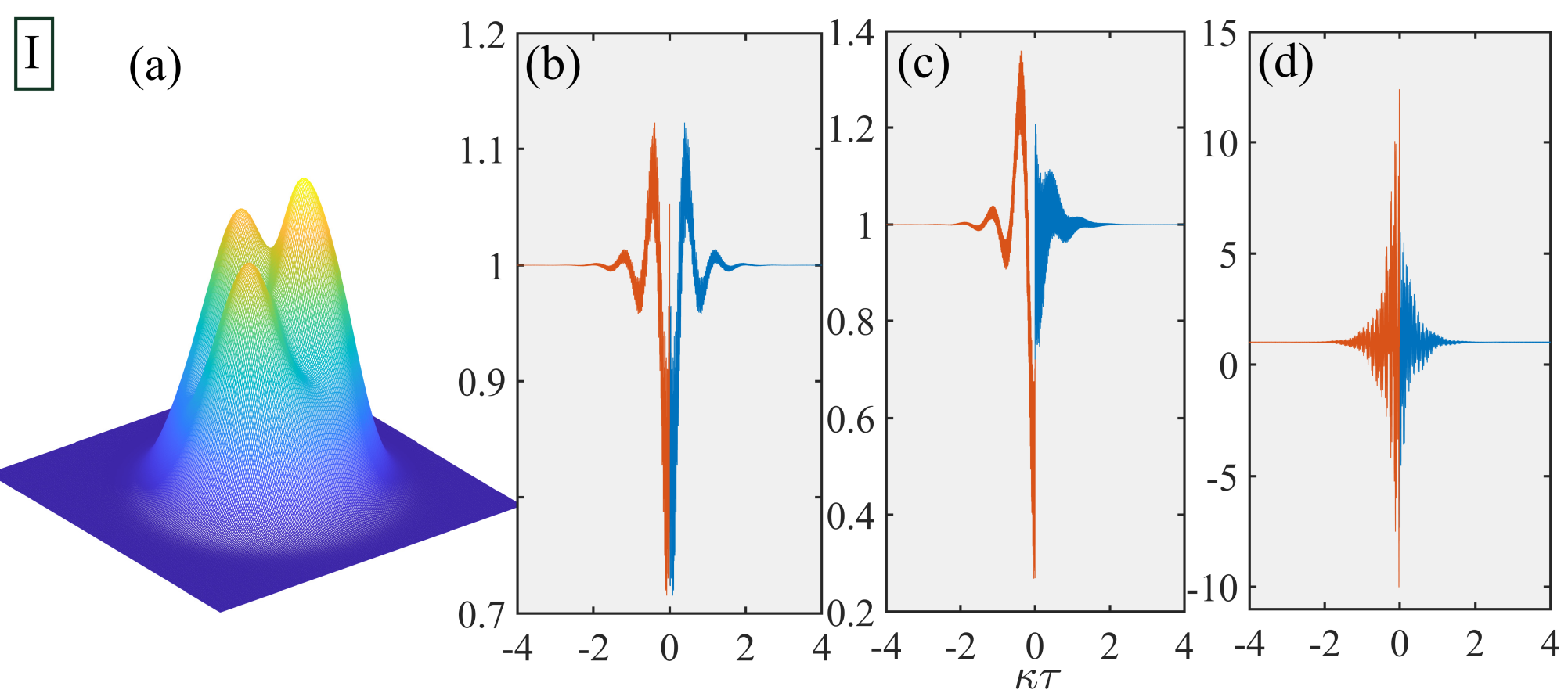}
  \includegraphics[width=\textwidth]{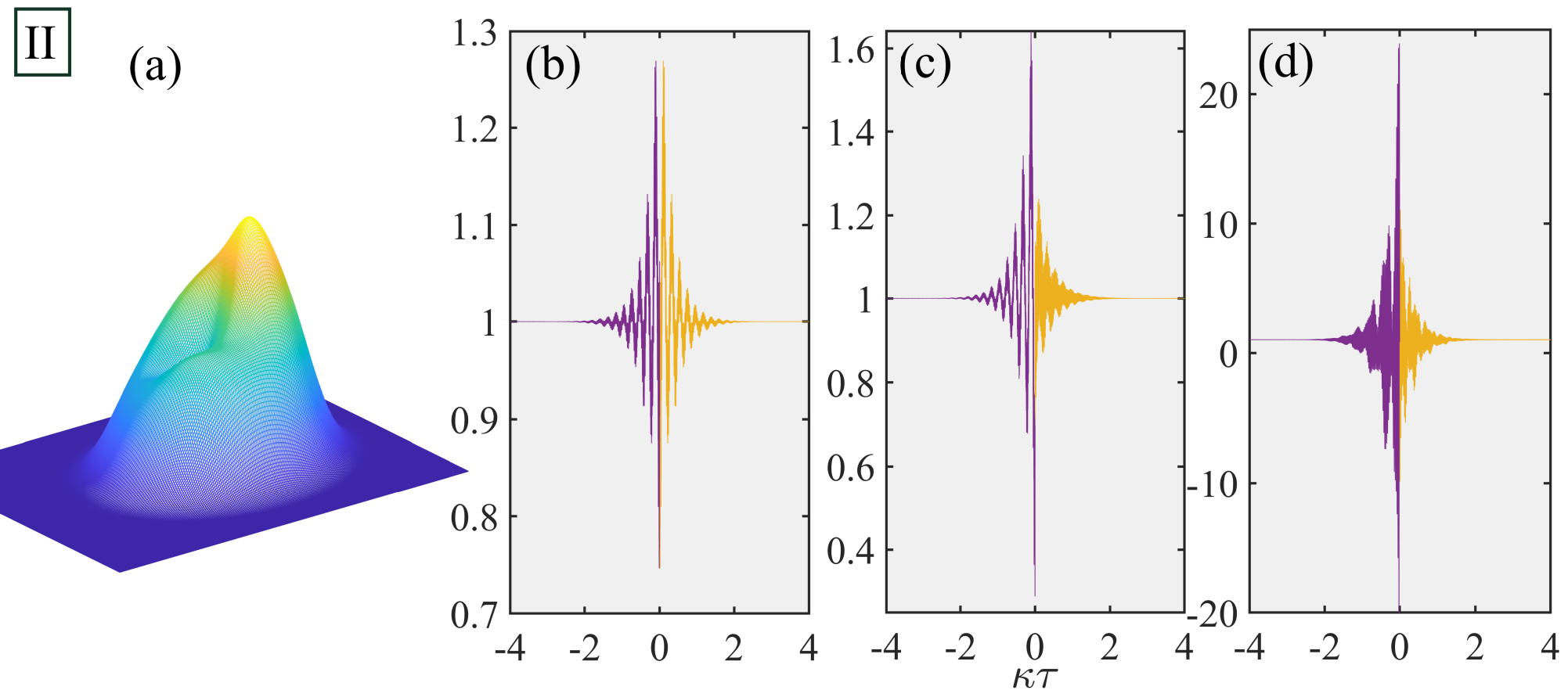}
   \includegraphics[width=\textwidth]{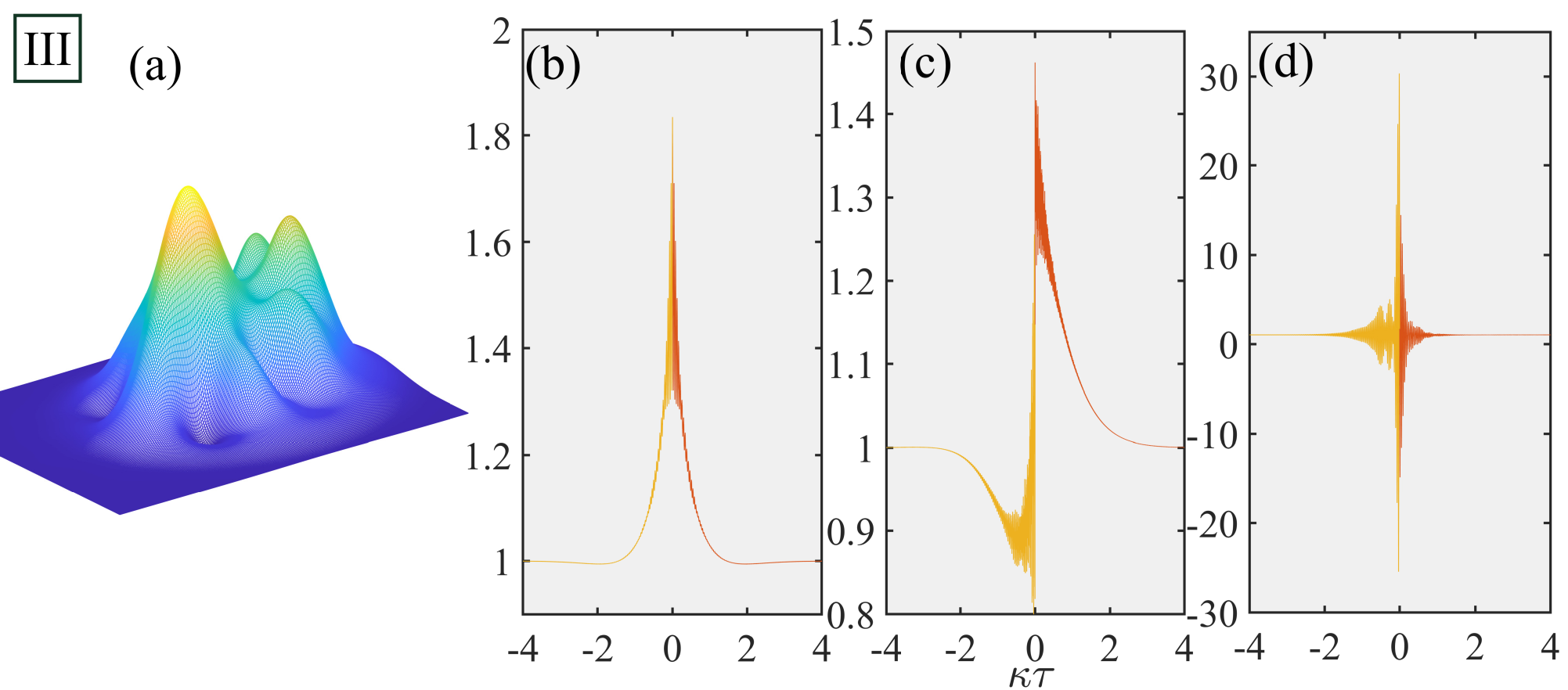}
\end{figure*}

\begin{figure*}
  \caption{{\it Asymmetric fluctuations and breakdown of detailed balance.} Steady-state Wigner function $W_{\rm ss}(x+iy, x-iy)$ [frames {\bf (a)}], intensity correlation function of the forwards-scattered field $g^{(2)}(\tau)$ [frames {\bf (b)}], cross-correlation of the side and forward-scattered light intensity $g_{AB}^{(2)}(\tau)$ [frames {\bf (c)}] and normalized wave-particle correlation function $\overline{H}_{\theta}(\tau)$ [frames {\bf (d)}] with $\theta=\pi/2$. All these quantities are extracted from the numerical solution of the ME and the application of the quantum regression formula for the following operating points in the phase space ($\varepsilon_d/g, \Delta\omega_d/g$): $[(0.075, 0.5824), (0.12, 0.5903), (0.14, 0.38674)]$ in \underline{Panels I, II, III}, respectively. The difference in colors distinguishes negative ($\tau <0$) from positive ($\tau \geq 0$) time delays in all correlations.}
    \label{fig:fig3}
\end{figure*}

\begin{figure*}
\centering
 \includegraphics[width=\textwidth]{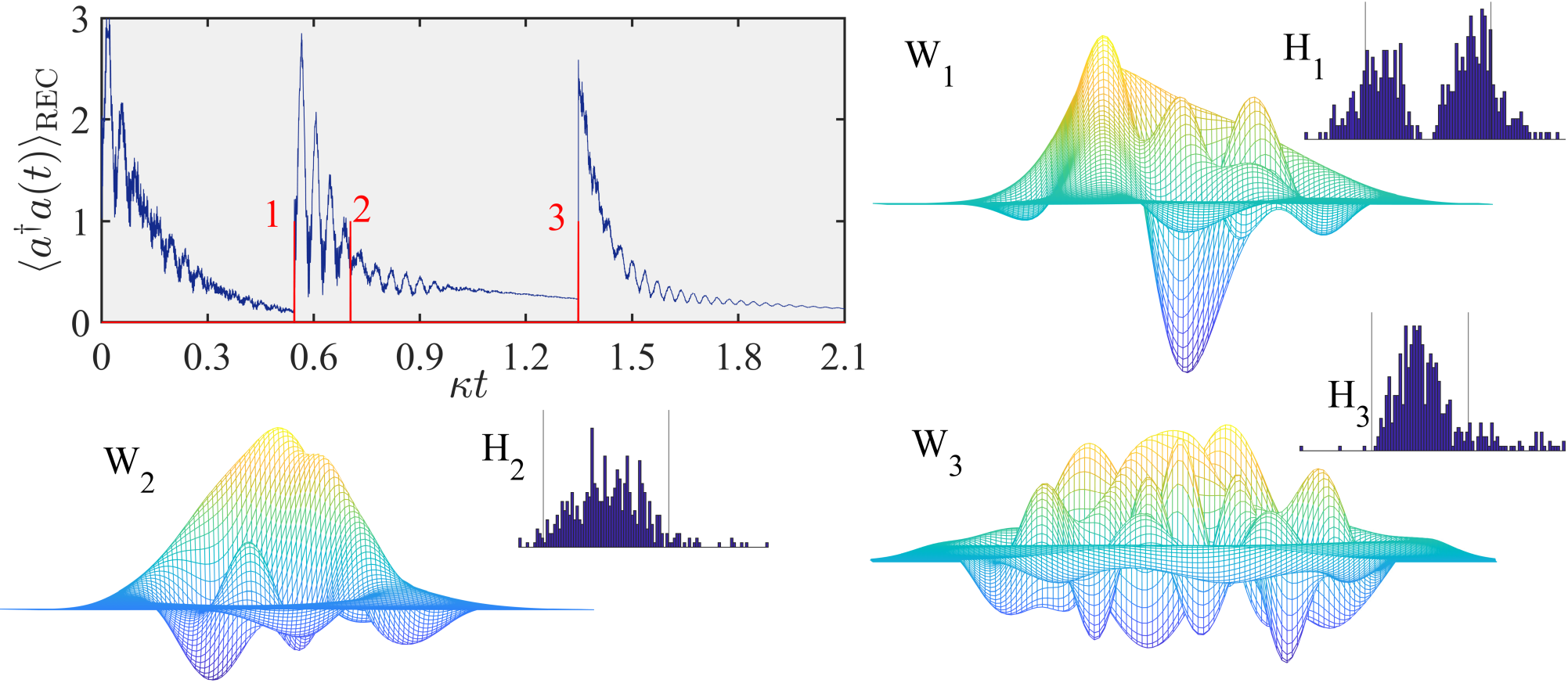}
 \caption{{\it Homodyne-detection tomography of photon bunching.}Fluctuating conditional photon number $\langle a^{\dagger}a (t) \rangle_{\rm REC}$ in the transient of a direct-photodetection unraveling of the seven-photon resonance. Three spontaneous emissions are indicated by the red strokes $1 ,2, 3$, conditioning the cavity field to the states with the Wigner functions (unlabeled surface plots) $W_1, W_2, W_3$, respectively. The corresponding histograms $H_1, H_2, H_3$ measure marginals of the Wigner function, $P_{\theta}(Q)$, along the direction $\theta=0$, extracted from an empty cavity measured for $\kappa T=7$ with homodyne detection. The left (right) vertical lines indicate the cumulative charge values $-1$ ($+1$), respectively. The remaining parameters read: $\varepsilon_d/g=0.38674$, $g/\kappa=10^3$, and the initial state is set to two-photon Fock state (with the two-level atom in the ground).}
 \label{fig:fig4}
\end{figure*} 

\begin{figure*}
\centering
 \includegraphics[width=\textwidth]{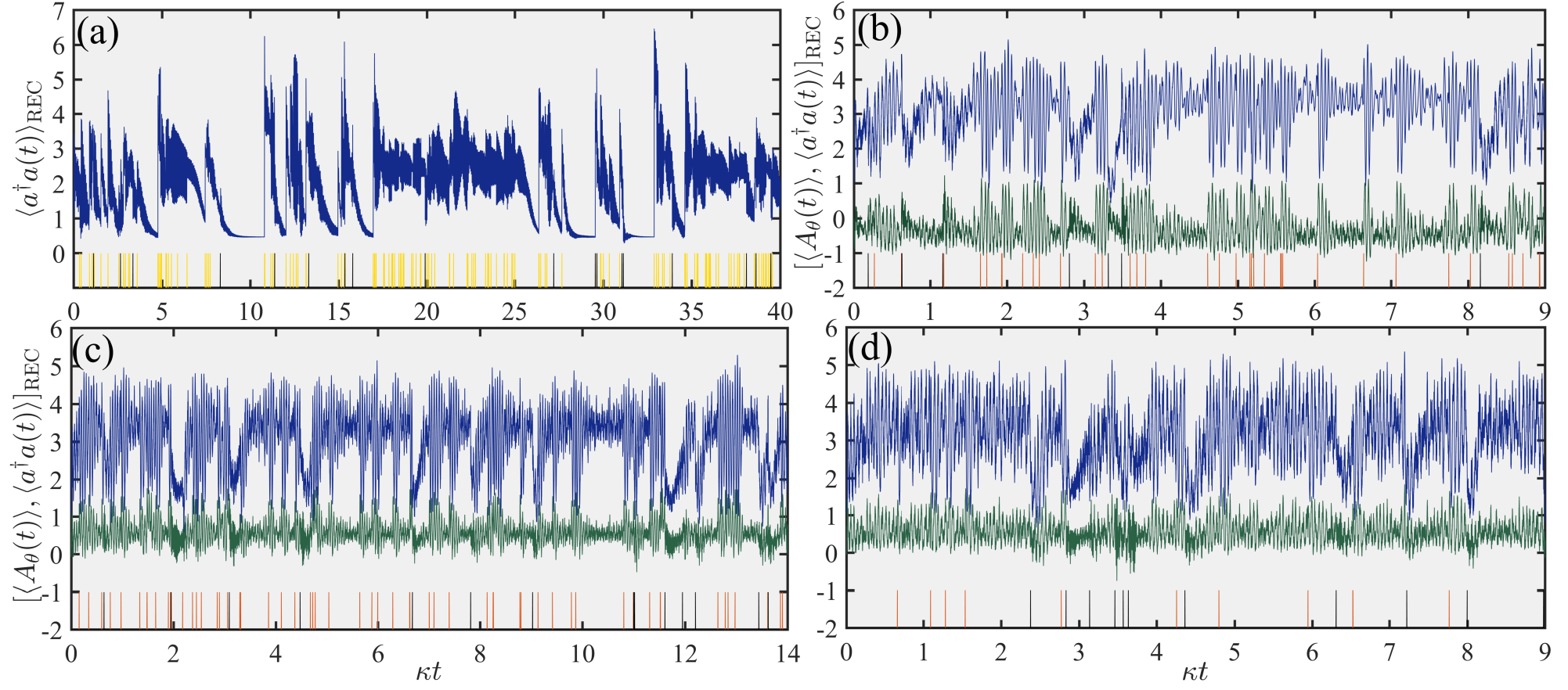}
 \caption{{\it Unraveling the seven-photon resonance peak.} Conditional photon number $\langle a^{\dagger}a (t) \rangle_{\rm REC}$ in blue (positive valued) and conditional fluctuating field $\langle A_{\theta}(t) \rangle_{\rm REC}$ in green (with alternating sign) for a direct photodetection unraveling in {\bf (a)} and three wave-particle unravelings in frames {\bf (b--d)}, against the dimensionless time $\kappa t$. The operating points $(r, \theta)$ read: $[(0.75, \pi/2), (0.75,0), (0.25,0)]$ in (b, c, d) respectively. Black strokes indicate spontaneous emissions. The operating parameters read: $\varepsilon_d/g=0.38674$, $g/\kappa=1,000$, and the initial state is set to two-photon Fock state (with the two-level atom in the ground state).}
 \label{fig:fig5}
\end{figure*}

\begin{figure*}
\centering
 \includegraphics[width=\textwidth]{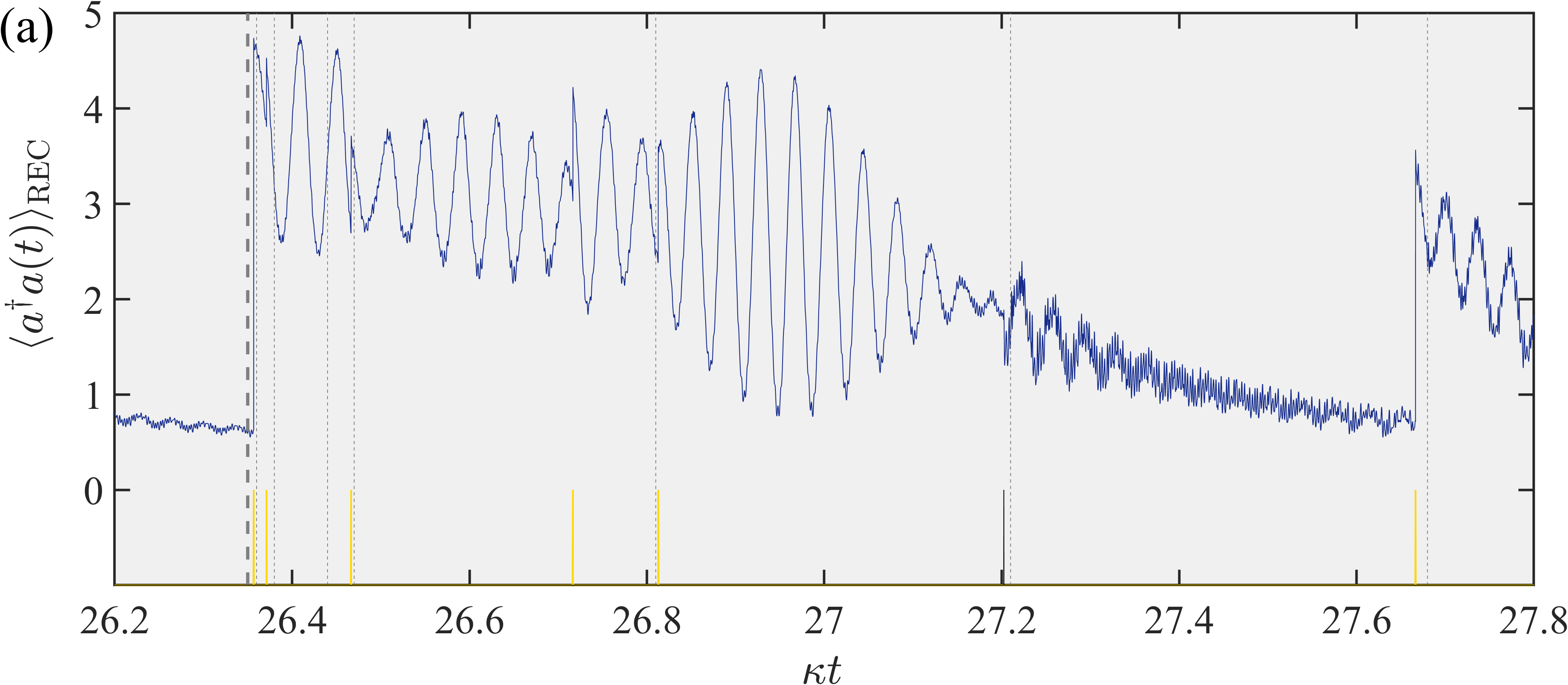}
  \includegraphics[width=\textwidth]{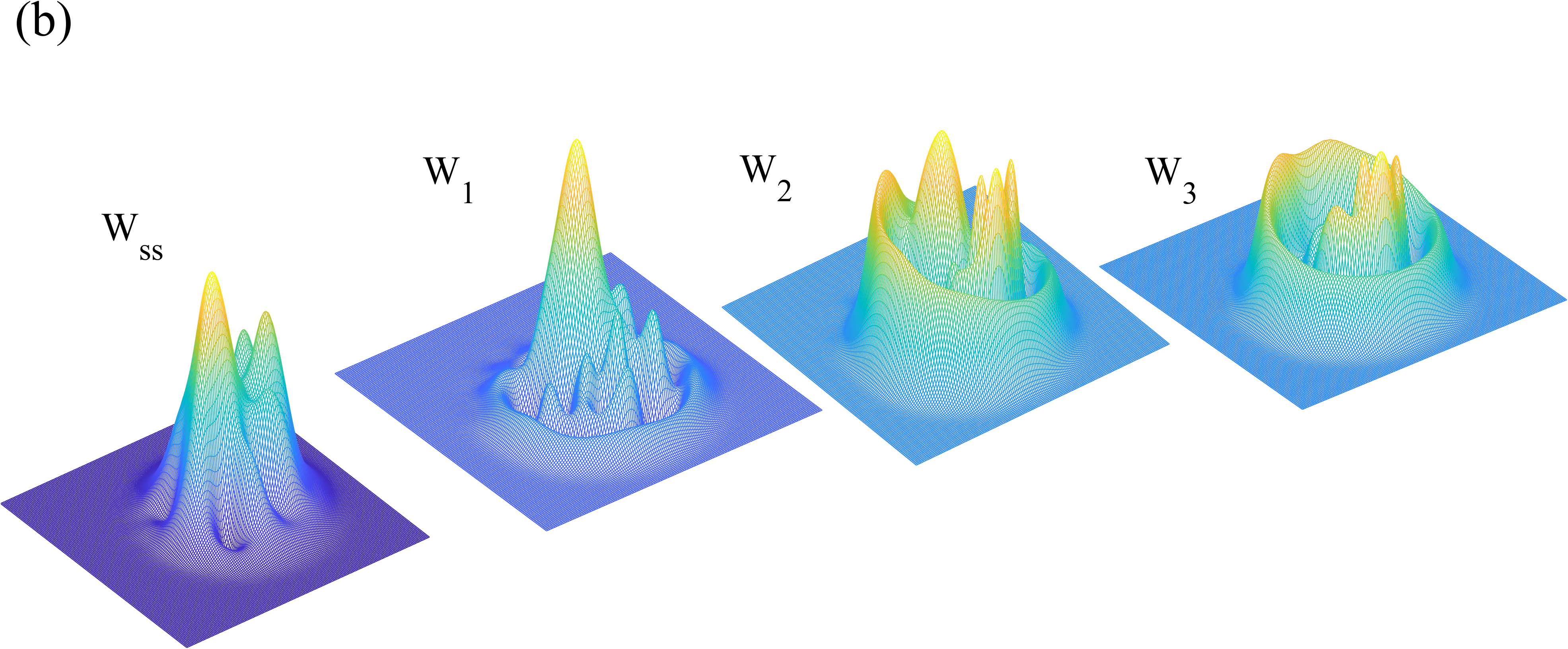}
   \includegraphics[width=\textwidth]{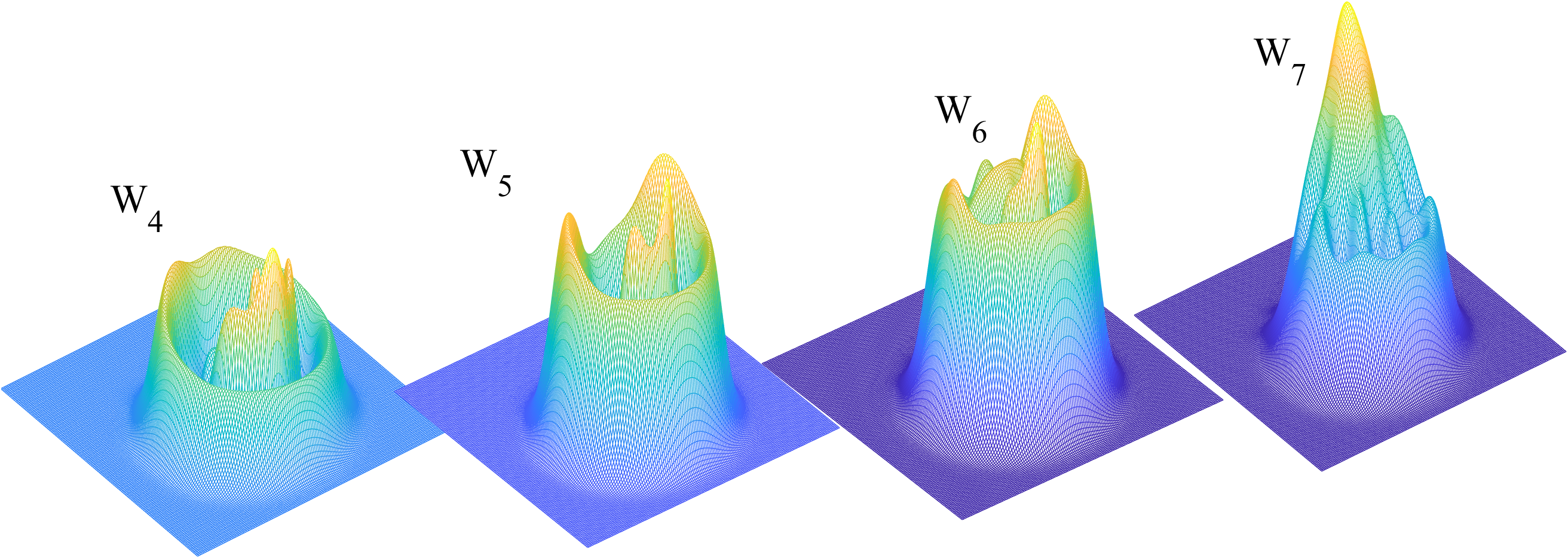}
   \caption{{\it Averaging over the density matrix across a cascaded decay.} {\bf (a)} A focus on a fragment of the time evolution under the direct photodetection unraveling of the seven-photon resonance given in Fig.~\ref{fig:fig5}(a), where a sequence of seven photons are emitted, one sideways and six in the forwards direction. {\bf (b)} Wigner functions (unlabeled surface plots) of the steady state ($W_{\rm ss}$) and of the time evolving density operator ($W_{1}-W_{7}$) of the cavity field averaged over seven time intervals, all starting at the time indicated by the thick dashed vertical line, and each ending at the sequential times indicated by the thin grey vertical lines, to conclude with an interval containing all seven emissions.}
   \label{fig:fig6}
\end{figure*}

\begin{figure*}
\centering
 \includegraphics[width=\textwidth]{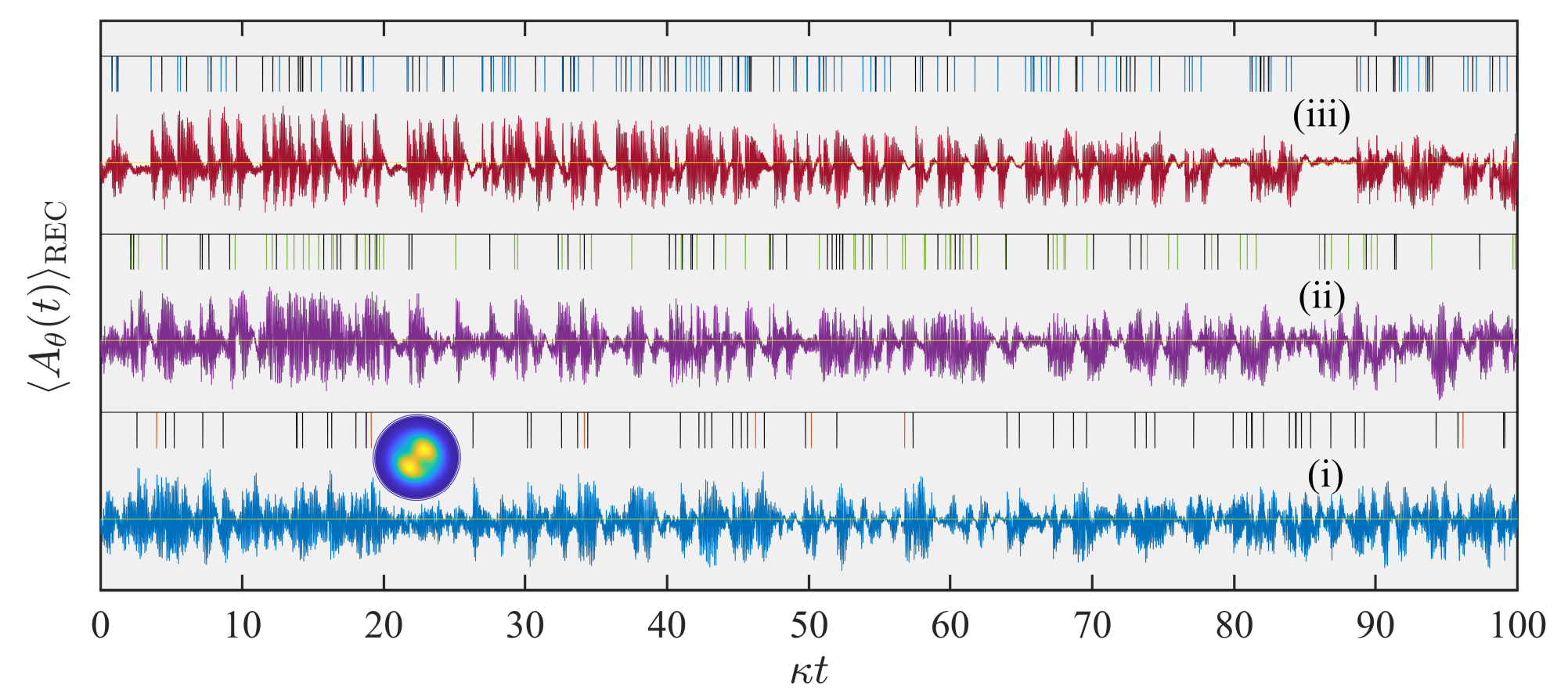}
  \caption{{\it Sweeping across the quadratures.} Fluctuating conditioned field amplitude $\langle A_{\theta}(t) \rangle_{\rm REC}$ for a time varying phase $\theta$ of the local oscillator, uniformly scanning the interval $[\pi/4, 3\pi/4]$ in the dimensionless time depicted. The branching ratio reads $r=0.05, 0.5, 0.95$ for the curves (i, ii, iii). The inset at the bottom left depicts a schematic contour of the steady-state Wigner function. Spontaneous emissions are indicated by black strokes, totaling $57, 52, 67$ side counts in the unravelings (i, ii, iii), respectively. The operating parameters read: $\varepsilon_d/g=0.08$, $g/\kappa=200$, and the initial state is set to $|0,-\rangle$.} 
  \label{fig:fig7}
\end{figure*}

\begin{figure*}
\centering
 \includegraphics[width=\textwidth]{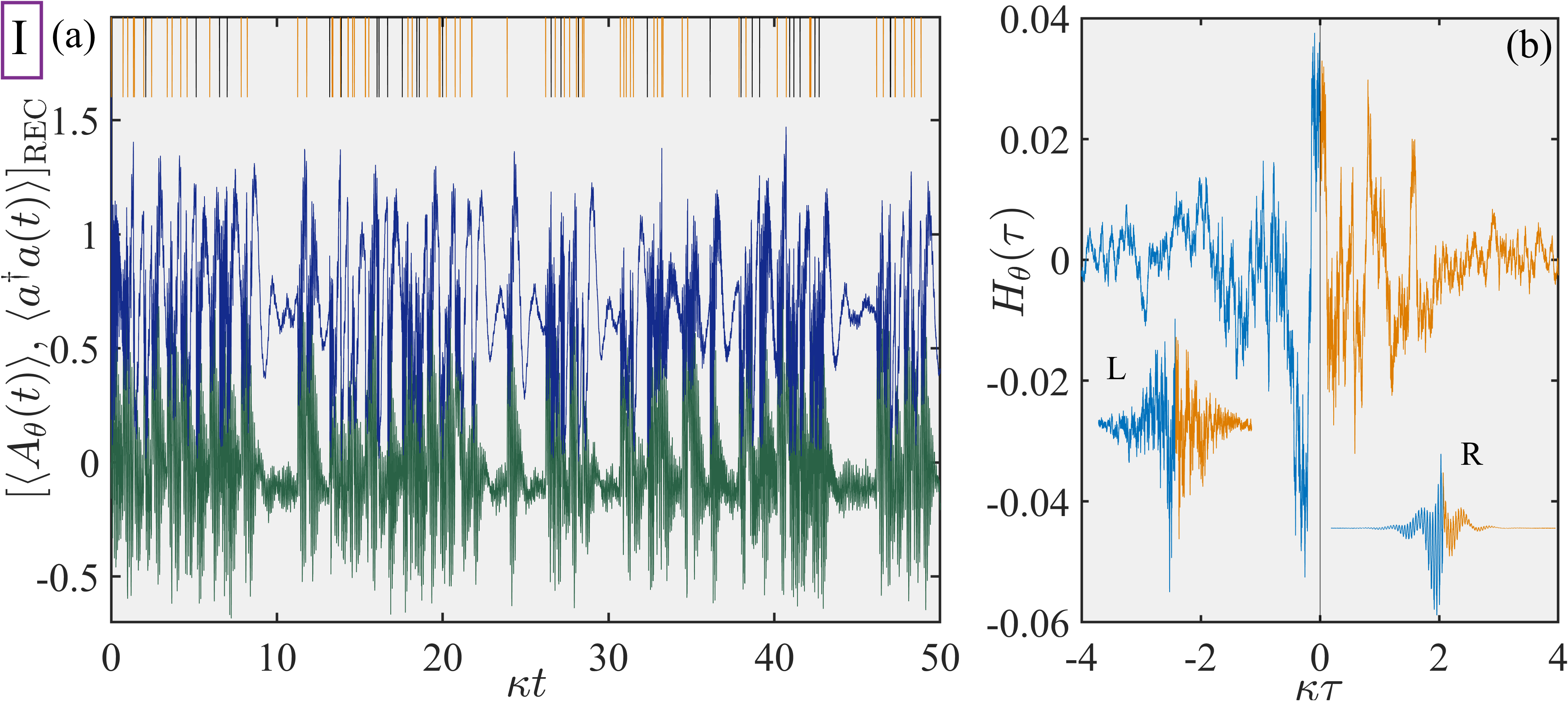}
  \includegraphics[width=\textwidth]{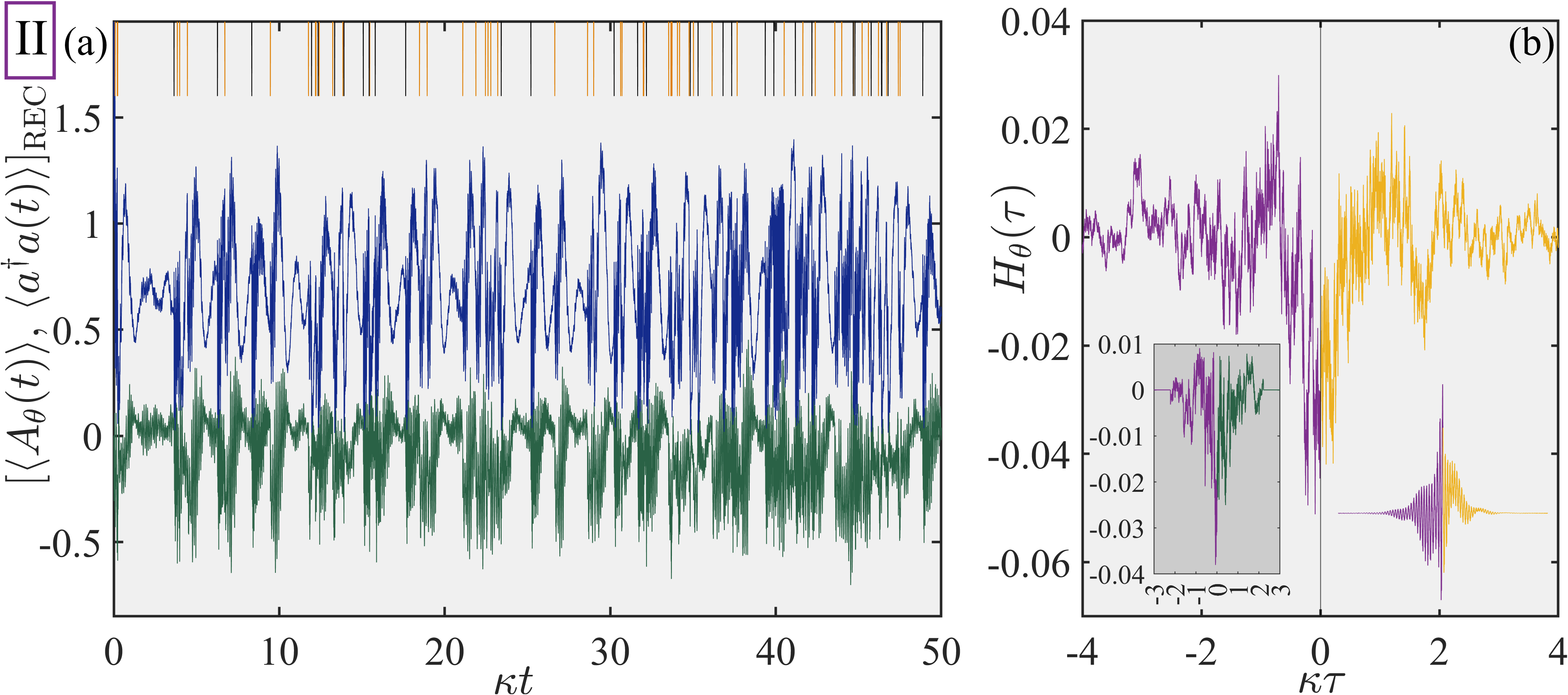}
   \includegraphics[width=\textwidth]{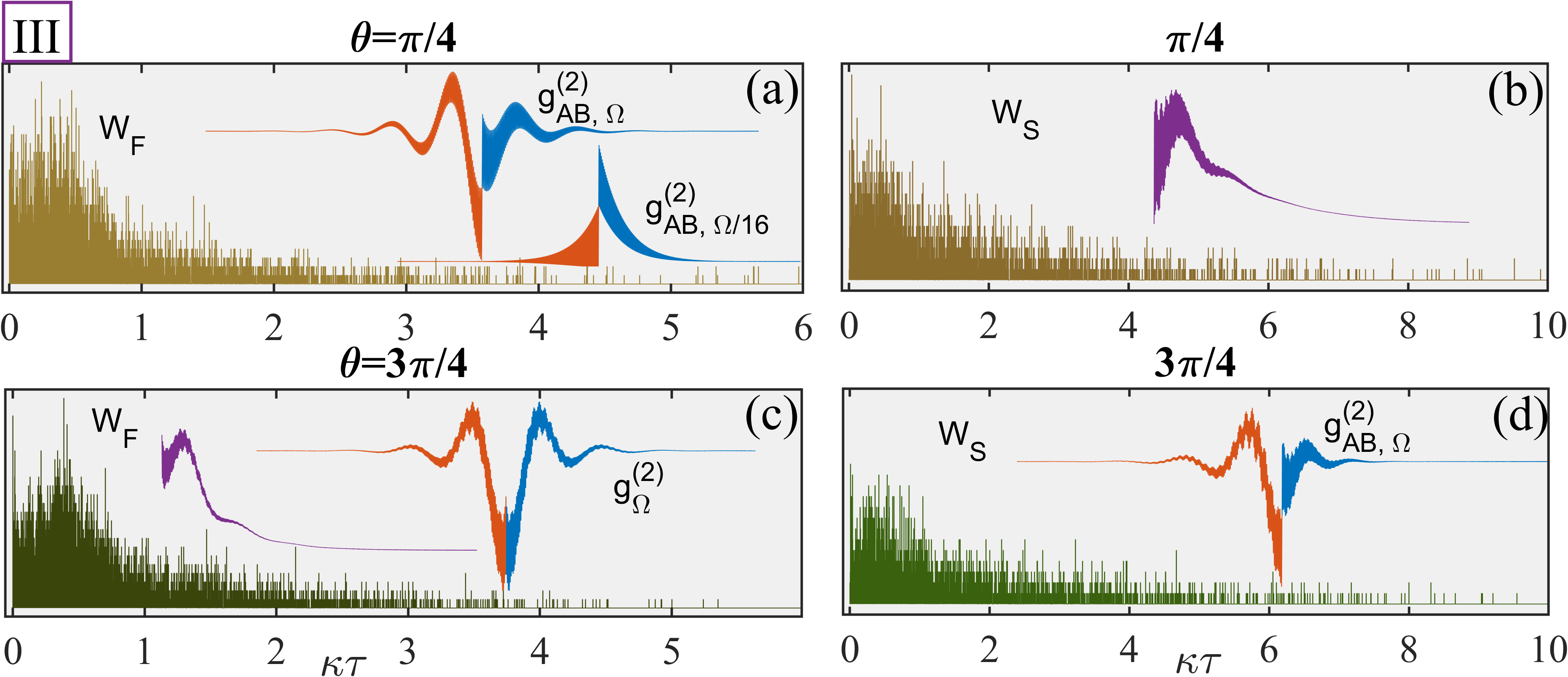}
\end{figure*} 

\begin{figure*}
    \includegraphics[width=\textwidth]{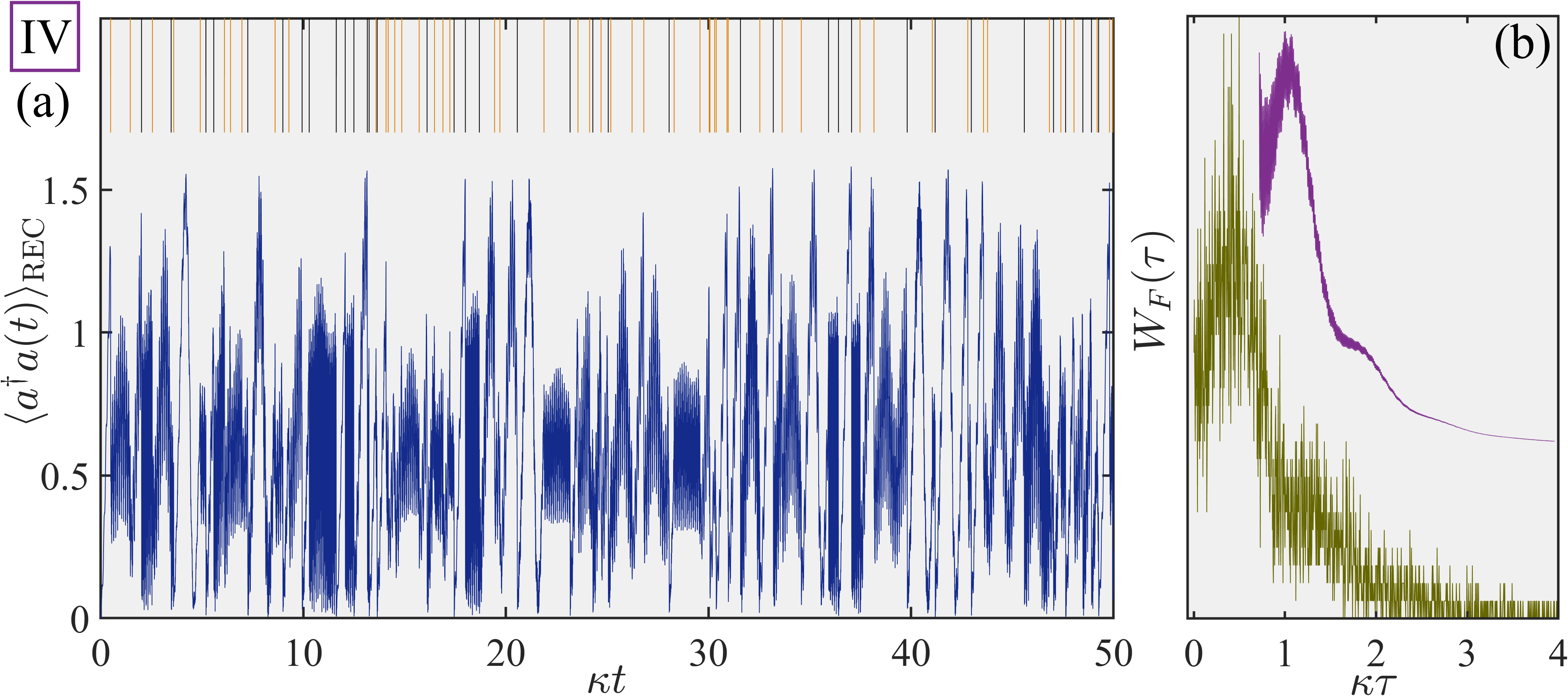}
  \caption{{\it Averaged photocurrents and photoelectron waiting times from single runs.} Fluctuating conditioned photon number (positive valued, in blue) and field amplitude (with alternating signs, in green) for the wave-particle unraveling of the two-photon resonance with $r=0.95$ in frames {\bf (a)} of \underline{Panels I and II}. In Panel I, the phase of the local oscillator is set to $\theta=\pi/4$, along the direction of bimodality, while Panel II probes the orthogonal direction. Frames {\bf (b)} depict the operationally determined wave-particle correlation function $H_{\theta}(\tau)$ extracted from similar individual runs but with $r=0.5$, numbering $31$ ($30$) cavity (spontaneous) emissions for Panel I and $24$ ($29$) cavity (spontaneous) emissions for Panel II, all for a detection bandwidth $B=10\kappa$. The two insets in the bottom right (R) depict unlabeled plots of the normalized numerically extracted $H_{\theta}(\tau)$ against four cavity lifetimes on either side (the different colors emphasize the asymmetry in time). The bottom left (L) inset of frame I (b) depicts the operationally determined wave-particle correlation for $169$ total clicks and detection bandwidth $B=30\kappa$. The bottom left (L) inset of frame II (b) presents the operationally determined $H_{\theta}(\tau)$ at $\theta=3\pi/4$ for $r=0.96$. The frames of \underline{Panel III} show photoelectron waiting time distributions for forwards [{\bf (a), (c)}] ($7057$ and $6885$ counts, respectively) and side [{\bf (b), (d)}] ($3417$ and $3319$ counts, respectively)  scattering in the form of histograms deriving from several individual runs, setting $\theta=\pi/4$ [{\bf (a), (b)}] and $\theta=3\pi/4$ [{\bf (c), (d)}]. The two inset curves in (a) plot  $g_{AB}^{(2)}(\tau)$ across eight cavity lifetimes, from Eq.~\eqref{eq:crosscorr} for $\varepsilon_d/g=0.08$ (top) and $\varepsilon_d/g=0.02$ (bottom). The two insets in purple in frames (b), (c) plot the waiting-time distributions $W_{S}(\tau)$, $W_{F}(\tau)$, respectively, extracted numerically from the ME and the quantum regression formula. The right inset of frame (c) and the inset of (d) give the numerically obtained intensity correlation function $g^{(2)}(\tau)$ and $g_{AB}^{(2)}(\tau)$, respectively, for $\varepsilon_d/g=0.08$. \underline{Panel IV} depicts a record obtained from photoelectron counting {\bf (a)} next to the photoelectron waiting-time distribution of the forwards scattered light obtained from $6062$ counts {\bf (b)}. The inset in (b) depicts the numerically obtained $W_F(\tau)$. The operating parameters read: $\varepsilon_d/g=0.08$, $\Delta\omega_d/g=1/\sqrt{2}+\sqrt{2}(\varepsilon_d/g)^2$, $g/\kappa=200$, and the initial state is everywhere set to $|0,-\rangle$.}
       \label{fig:fig8}
\end{figure*}

\begin{figure*}
\centering
 \includegraphics[width=\textwidth]{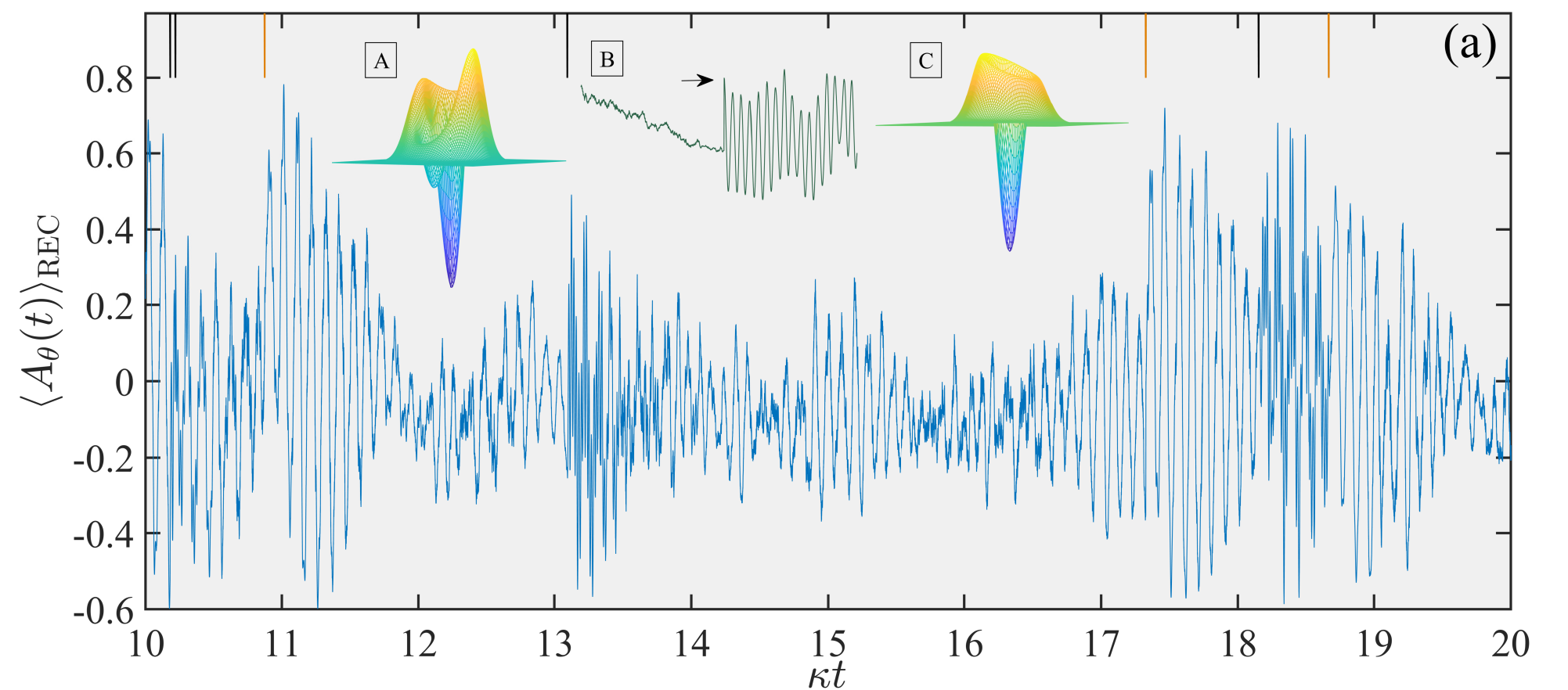}
  \includegraphics[width=\textwidth]{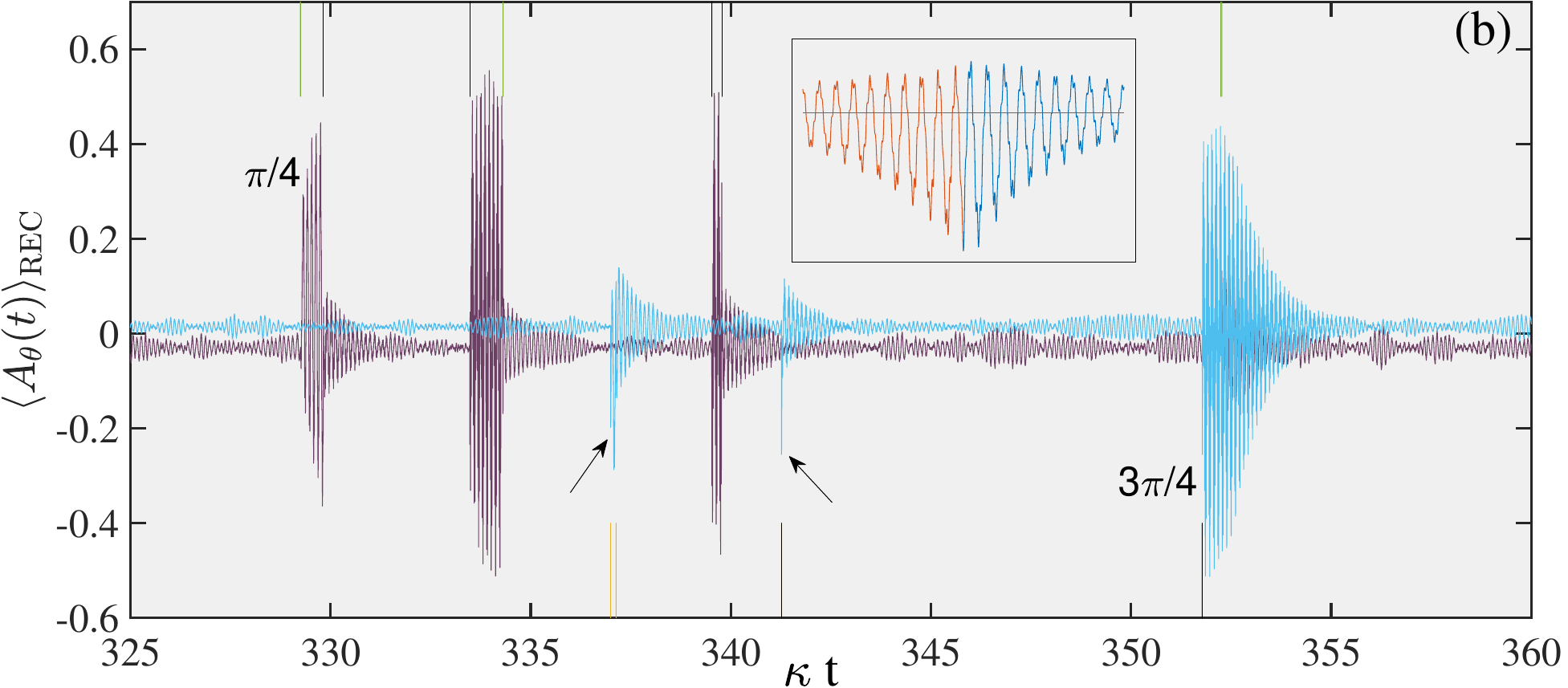}
   \includegraphics[width=\textwidth]{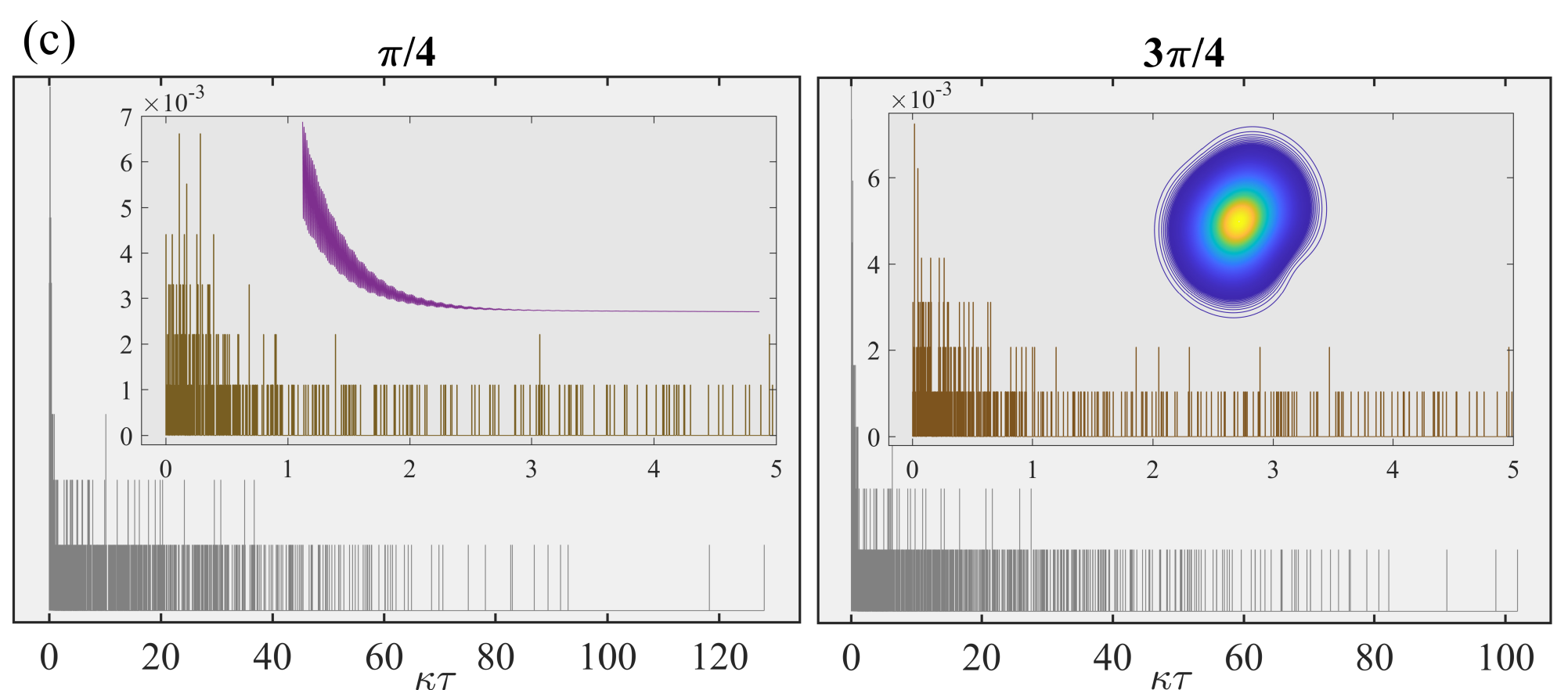}
\end{figure*}

\begin{figure*}
   \caption{{\it Field fluctuations along the development of bimodality.} {\bf (a)} Fluctuating conditioned field amplitude $\langle A_{\theta} \rangle_{\rm REC}$ against the dimensionless time $\kappa t$ in a fragment of a single realization, at $\theta=\pi/4$ for an equally balanced ($r=0.5$) wave-particle unraveling of the two-photon resonance ME and for $\varepsilon_d/g=0.08$. Spontaneous (cavity) emissions are indicated by black (orange) strokes. The three insets correspond to a transient similar to what is shown in Fig.~\ref{fig:fig4}. Inset A depicts the Wigner function (surface plot) of the cavity field at the time when the maximum conditional photon number $\langle a^{\dagger}a \rangle_{\rm REC, max}\approx 1.37$ is attained, inset B shows the revival of the quantum beat in the fluctuating photon number following a spontaneous emission (conditioning the cavity in a state with $\langle a^{\dagger}a \rangle_{\rm REC}=1$), and inset C depicts the Wigner function (surface plot) at the latter quantum jump, marked by the arrow in B. {\bf (b)} Same as in (a) but for $\varepsilon_d/g=0.02$ and two values of $\theta$: $\pi/4, 3\pi/4$ as indicated next to each curve. Spontaneous (cavity) emissions are indicated by black (green--$\pi/4$, orange--$3\pi/4$) strokes. The framed inset depicts the normalized numerically extracted $H_{\theta=3\pi/4}(\tau)$ (the black line marks the zero level) across one cavity lifetime on either side, while the arrows point to the anomalous phase. {\bf (c)} Waiting-time distributions of the forwards-scattered light for $\varepsilon_d/g=0.02$ and two different settings of $\theta$, indicated on top of each frame ($906$ and $965$ counts for $\theta=\pi/4, 3\pi/4$, respectively). The two insets focus on the initial waiting times, while the unlabeled purple curve is extracted from the ME (for the same times as in the inset) and the contour plot depicts the steady-state Wigner distribution. The operating parameters read: $\Delta\omega_d/g=1/\sqrt{2}+\sqrt{2}(\varepsilon_d/g)^2$, $g/\kappa=200$.}
     \label{fig:fig9}
\end{figure*}

Panel II of Fig.~\ref{fig:fig2} focuses on the dynamical consequences of alternating bunching and antibunching as the cavity detuning traverses the ``vacuum'' Rabi, two- and three-photon resonances. Upon comparing the photon emissions about the ``vacuum'' Rabi resonance (which is predictably strongly anti-bunched) and the two-photon resonance we find a triplet-pattern repeating: in the former (at $|\Delta\omega_d/g| \approx 0.95$) the triplet contains single counts, while in the latter (at $|\Delta\omega_d/g| \approx 0.7$) we observe a triplet of photon pairs, each of which corresponds to a side and a forwards emission. In anticipating the three-photon resonance with a bunched output, the same number of photon emissions (six) are now rearranged into two triads, keeping the balance between spontaneous and cavity emissions. Perhaps a more insightful glimpse at the dynamics is allowed by the more refined time discretization in inset L, focused on the two-photon resonance peak. A closely-spaced double emission at about $|\Delta\omega_d/g| \approx 0.716$ (not resolved on the scale of the figure) initiates the semiclassical ringing~\cite{Shamailov2010, Mavrogordatos2024} (and the superimposed quantum beat), while in the direction of smaller detunings the conditioned photon number appears to closely follow the steady-state two-photon resonance peak albeit with an even greater amplitude, exceeding unity. The ascent is interrupted by a cavity emission -- a characteristic sign of photon antibunching in a very narrow range of detuning values about $|\Delta\omega_d/g| \approx 0.71$ (see inset C in the middle). The two intense re-initiations of the quantum beat are concluded by a spontaneous emission, which brings the semiclassical oscillation back on track to `conclude' the second-photon resonance out. Side emissions slightly exceed cavity jumps. 

A five-fold increase of the drive amplitude in Panel III of Fig.~\ref{fig:fig2} significantly saturates the one, two and three-photon transitions, while the zero-delay correlation approaches unity from both the extreme bunching and antibunching ends; its value is actually brought closer to the steady-state photon occupation but with anti-correlated dips and peaks. The steady-state photon number is recovered as a quasi-continuous dynamical average of a quantum beat (or an interference of quantum beats about the three-photon resonance peak) in what makes the conditioned probability of emission, interrupted only some occasions by instances of photon bunching or antibunching. Now the jumps are rearranged from pairs to triads, while the number of cavity emissions exceeds the double of side-emission ``clicks''.

In contrast to Fig.~\ref{fig:fig2}, Fig.~\ref{fig:fig3} presents results solely obtained from the ME and the quantum regression formula. The plots focus on the characteristic nonlinearity and the timescales in the dynamics associated with a three- and a seven-photon transition in conjunction with the temporal asymmetry of quantum fluctuations. Panels I and II illustrate the change in the statistics of light -- in its dual character -- accompanying the saturation of the three-photon transition. Both operation settings yield the same zero-delay autocorrelation, $g^{(2)}(0) \approx 1.05$. Nevertheless, the phase-space distribution of the cavity field evinces a reorganization of steady-state quantum fluctuations towards the collapse of tri-modality. Further testifying to the breakdown of detailed balance, negative time delays display a pronounced peak and a dip in the cross-correlation between side and forwards emissions, as well as more intense amplitude-intensity correlations whose maximum peak at $\theta=\pi/2$. In Panel III, the multi-modality associated with the peak of the seven-photon transition underlies now very different particle-particle correlations: photon bunching is more pronounced, while the intermediate timescale (between the quantum beat and the semiclassical ringing) we encountered in the saturation of the three-photon resonance is now unresolved. Nevertheless, the temporal asymmetry of quantum fluctuations persists. Let us now complement the ME results with individual realizations conditioned on a particular series of detector ``clicks'' in various settings of the wave-particle correlator. 

\subsection{Operational focus on the seven-photon resonance peak}

Marking a departure from the dual function of the wave-particle correlator for the time being, let us concentrate on direct photodetection {\it followed} by homodyne/heterodyne detection to operationally infer the {\it quasi} probability distribution of the cavity field conditioned on a given quantum jump or a series of jumps. In Fig.~\ref{fig:fig4}, a photoelectron counting record is taken for a set time $T$ in the transient, somewhat exceeding the average photon lifetime from a cavity driven to host a seven-photon resonance. We observe the gradual loss of the circular symmetry in the Wigner function of the initially prepared two-photon Fock state, in a background of three spontaneous emission events and photon bunching [see Fig.~\ref{fig:fig3}III(b)]. Let us suppose that after any one of the three spontaneous emission events, we might have selected to turn off the drive ($\varepsilon_d=0$) and light-matter coupling strength ($g=0$), allowing the cavity field to freely decay into the modes of the vacuum. In the course of the decay, the record keeping is switched to homodyne detection though Path I in Fig.~\ref{fig:fig1} with $r=0$. The local oscillator phase is set to $\theta=0$ (probing the $X$-quadrature of the field), while the local oscillator amplitude is also exponentially decaying, temporally mode-matched to the cavity field. Throughout the decay, the net charge deposited into the detector is integrated over time, such that when all the light has left the cavity, a cumulative charge has been produced (a real number), reading
\begin{equation}\label{eq:cumulativecharge}
 Q=\sqrt{\kappa/2}(Ge |\varepsilon_{\rm lo}|)^{-1}\int_{T}^{\infty}e^{-\kappa(t-T)}dq,
\end{equation}
where $G$ is the detector gain and $e$ is the electronic charge. Defining $\langle A_{\theta} \rangle_{\rm REC}\equiv (1/2)\langle e^{i\theta} a^{\dagger} +  e^{-i\theta} a\rangle_{\rm REC}$ the conditioned quadrature amplitude selected by the local-oscillator phase, the elemental charge deposited in the detector is 
\begin{equation}
 dq=Ge |\varepsilon_{\rm lo}| \sqrt{2\kappa}\left(2\langle A_{\theta} \rangle_{\rm REC}\,dt + dW \right),
\end{equation}
and the conditioned (un-normalized) state obeys the stochastic Schr\"{o}dinger equation~\cite{CarmichaelQO2}
\begin{equation}
\begin{aligned}
 d|\psi_{\rm REC}\rangle=&[-i\Delta\omega_d a^{\dagger}a\,dt-\kappa a^{\dagger}a\,dt \\
 &+ (Ge |\varepsilon_{\rm lo}|)^{-1}e^{-i\theta}\sqrt{2\kappa}a\,dq] |\psi_{\rm REC}\rangle.
 \end{aligned}
\end{equation}
The probability distribution $P_{\theta}(Q)$ thus obtained measures a marginal of the Wigner {\it quasi}probability distribution representing the state of the intracavity field immediately before the time interval of the free decay~\cite{CarmichaelKochan1994, CarmichaelQO2}. Instead of separating the two schemes in time -- direct photon counting followed by homodyning -- let us now merge them.  

Figure~\ref{fig:fig5} focuses on complementary unravelings of the ME when operating at the peak of the seven-photon resonance. In these unravelings, an externally imposed asymmetry between the forwards and side emissions is tuned by hand through an imbalance of fluxes directed towards the two arms of the correlator (we still operate with $\gamma=2\kappa$). The direct photodetection in frame (a) evinces recurring relaxations at a low-photon state, below the steady-state value $\langle a^{\dagger}a\rangle_{\rm ss} \approx 1.83$. The relaxation is interrupted by peaks in the conditioned photon number whenever a quantum jump is recorded. Here the cavity emissions massively overpopulate the side emissions by a larger factor than the one we found in Panel III of Fig.~\ref{fig:fig2}. Photon emission is bunched with $g^{(2)}(0) \approx 1.83$, and the number of cavity emissions massively outnumbers spontaneous emission events despite the fact that $\gamma=2\kappa$. Meanwhile, a set of metastable states are visited when long bunched photon sequences are being recorded, e.g, for $20 \leq \kappa t \leq 25$, together with an interference of quantum beats. 

A rather different picture is offered by the wave-particle correlator unraveling associated with monitoring a set quadrature of the intracavity field, recorded in frames (b-d). On the one hand, the field is found in a high photon state between the trigger jumps, while cavity emissions induce rapid high-amplitude fluctuations. On the other hand, spontaneous emissions bring the cavity to a low-photon state. This becomes clear in frame (d) where $r=0.25$ and the number of forwards jumps equals that of side jumps during the first nine cavity lifetimes. Comparing frames (b) and (c,d) we find that between the jumps, the conditioned field relaxes to a negative (positive) value at $\theta=\pi/2$ ($\theta=0$). For both values of $\theta$, forwards or side jumps do not change the sign of the conditioned field, as reflected in the positive sign of the normalized $H_{\theta}(0)$.

In Fig.~\ref{fig:fig6}, we bring up a particular aspect of wave-particle duality by relating the cascaded process underlying the seven-photon resonance to the averaged value of the cavity-field distribution after sequential emissions are being recorded. To that aim, we revisit frame (a) of Fig.~\ref{fig:fig5}. Six cavity and one spontaneous emissions are noted in a time shorter than the average photon lifetime. The latter instigates a reappearance of a superposition of quantum beats. The familiar pattern of photon antibunching [$g^{(2)}(0) \approx 1.83$] is created: after the first photon click from the bundle, the conditional emission probability $2\kappa\langle a^{\dagger}a(t)\rangle_{\rm REC}\,dt$ increases in anticipation of a second imminent cavity emission; remarkably, all cavity emissions depicted in Fig.~\ref{fig:fig6} follow this pattern. Only the sole spontaneous emission at $\kappa t_s \approx 27.2$ executes the awaited drop and brings the cavity into a coherent superposition of JC dressed states.

Let us now talk about the field distribution in the aftermath of the individual events substantiating the cascaded process. For a single trajectory continuously run in time, the steady-state system density matrix can be computed as an integral over a pure-state resolution,
\begin{equation}
 \rho_{\rm ss}=\frac{1}{T}\int_{0}^{T}dt\,|\psi_{\rm REC}(t)\rangle \langle \psi_{\rm REC}(t)|,
\end{equation}
where $T$ is much larger than the system decoherence time. The time average is equivalent to the ensemble average over records, since quantum trajectories are ergodic in the vast majority of cases~\cite{Cresser2001, CarmichaelQO2}. The equivalence prompts us to compare the steady-state distribution with the averaged contribution of successive jumps that make the cascaded process underlying a multiphoton resonance. In this example, we focus on one of the several seven-photon bundles selected from a single realization past the transient. The cascade of seven photon emissions is completed in a time shorter than the average photon lifetime.  

The density matrix starts being averaged from the time instant marked by the thick grey line at $\kappa t_{\rm start}=26.35$. After the first cavity emission, the multimodal structure in the Wigner function is undermined, while a de-excitation path develops encircling all peaks, similar to the ```skirt'' noted in~\cite{PhotonBlockade2015}, while the conditional distribution begins to turn negative in places. Longer averages record increasingly negative values corresponding to conditional interferences of Fock states, while the ``skirt'' joining the peaks is now raised higher. The last emission of the group generates a cavity-field distribution much closer to the steady-state profile, and the Wigner function remains everywhere positive. 

\subsection{Wave/particle duality at the two-photon resonance: conditioned field amplitudes and waiting-time distributions}

For the remainder of this report we concern ourselves with the quantum nonlinearity displayed in the measurement outcomes from the two-photon JC resonance, complementing the analytical results presented in Sec.~\ref{sec:particlecorsscorr}. Figure~\ref{fig:fig7} focuses on the impact of the rate at which conditional measurements of the field are being made. In all three records the local oscillator phase varies uniformly in time from $\theta=\pi/4$ to $\theta=3\pi/4$, while the fraction of the flux sent to the APD dramatically changes from $r=0.05$ (i) to $r=0.95$ (iii). The forwards-scattered light is antibunched, with the ME and the quantum regression formula giving $g^{(2)}(0) \approx 0.82$. It is only in the frequent sampling of the homodyne current that the conditioned field visibly drops to levels below zero as $\theta$ approaches $3\pi/4$, in line with the steady-state prediction of the ME. The drop is allowed by the extra cavity emissions triggering the sampling through the increased fraction $r$. All records evince bistable switching at the start of the scan, but the more frequent sampling distinguishes a clearer regression of fluctuations. It is worth concentrating on the time interval $84.04 < \kappa t < 88.7$ lapsing between a cavity emission and a spontaneous emission at the two ends in record (iii). The former, conditioning a field amplitude of the same sign, instigates a long relaxation about zero, while the latter conditions the field to a changing sign and marks an intense and distinct reappearance of the quantum beat. In the corresponding interval of record (i), occurring for $89.2<\kappa t<95.82$, there is no sign of fluctuation regression between the two spontaneous-emission events. In this extreme case there has been no sampling of the homodyne current for more than $30$ cavity lifetimes, which effectively means that there is no feedback to the conditioned wavefunction from clicks at the APD. Hence, there is no decay of fluctuations associated with a photocounting record making ($r\to 1$); decoherence via the spontaneous-emission channel does not induce a sufficient decay in the non-unitary Schr\"odinger equation. 

Carrying on with this theme, in Fig.~\ref{fig:fig8} we register the conditional evolution of the electromagnetic field when the wave-particle correlator operates from $r=0.5$ to $r \to 1$. We focus on the two orthogonal quadratures relevant to the display of bimodality. Frames (a) in Panels I and II have $r=0.95$, with $\theta$ set to $\pi/4$ and $3\pi/4$, respectively. The conditional photon number doesn't reveal any notable different between the two unravelings. In the absence of quantum jumps there is a relaxation close to the steady-state value $\langle a^{\dagger}a \rangle_{\rm ss} \approx 0.61$. Along the direction of bimodality, the conditioned field relaxes to a negative value, whence the positive steady state ($\langle A_{\theta=\pi/4} \rangle_{\rm ss} \approx 0.11$) is to be maintained by the occurrence of jumps interrupting those relaxations -- guaranteed by antibunching in the photon sequences [$g^{(2)}(0) \approx 0.82$]. While the number of spontaneous emissions remains the same in the trajectories of Panels I and II, a systematic 5\% decrease in the cavity emission events is noted when $\theta$ varies from $\pi/4$ to $3\pi/4$. Even this small correlation with the selected quadrature amplitude aligned with the onset of bimodality is a notable feature since, on the one hand, the steady-state cavity occupation barely exceeds half a photon and, on the other hand, a very small fraction of the cavity output is being directed to the BHD. 

The operationally determined wave-particle correlation function [see Eq.~\eqref{eq:HNs} and Appendix~\ref{sec:thewp}] of frames (b) in Panels I and II is determined from the sampled homodyne photocurrent for $r=0.5$, with a detection bandwidth ten times the cavity decay rate, $B=10\kappa$. (Normalizing by the negative steady-state field amplitude would mirror the two operationally obtained correlations in Fig.~\ref{fig:fig8}II(b) with respect to the time axis). The two insets at the bottom depict the normalized correlation over the same range, calculated from the ME and the quantum regression formula. The detection bandwidth cannot resolve the quantum beat in the strong-coupling limit (see App.~\ref{sec:thewp}). Nevertheless, a few tens of samples collected from single realizations capture more than the salient features of the third order correlation function. First and foremost, they evince the temporal asymmetry of quantum fluctuations for both settings of $\theta$. They recover the envelope of $H_{\theta}(\tau)$ obtained by the numerical solution of the ME, while occasional fluctuations identify the intermediate timescale (see App.~\ref{sec:4levelcross}) which dominates the wave-particle correlation. An increased detection bandwidth resolves the intermediate timescale.  A more frequent sampling [inset of Fig.~\ref{fig:fig8}II(b)] better resolves the time asymmetry about the origin by fails at longer delays.

Panel III informs us on the asymmetry (imbalance) of the times waited between successive cavity emissions when the local oscillator is aligned along and and orthogonal to the direction of bimodality, keeping $r=0.95$. The waiting-time distribution~\cite{Carmichael1989,Brandes2008} calculated from the ME and the quantum regression formula [left inset in (c)] predicts an average time between emissions equal to $\kappa \overline{\tau}\approx 0.82$, which is longer than the average time predicted by the distributions generated along the two quadratures. Specifically, we obtain $\kappa \overline{\tau}_{\pi/4}\approx 0.74$ and $\kappa \overline{\tau}_{3\pi/4}\approx 0.76$ along and orthogonal to the development of bimodality, respectively. Both distributions $W_F$ [frames (a) and (c)], in the form of histograms generated from about $7,000$ counts, display a rapid modulation due to the quantum beat, while the variance in the waiting times is 6\% smaller orthogonal vs. along the axis of bimodality. Surprisingly, a subtle yet systematic disparity is also observed between photoelectric counts recorded sideways for the two different unravelings: a smaller average time between emissions is noted at a direction orthogonal to bimodality. This panel is complemented by the (normalized) cross-correlation of intensity between forwards and side-emitted photons, analytically derived in Sec.~\ref{sec:particlecorsscorr} using the minimal four-level model, for the same drive amplitude ($\Omega$). Analytical [III, frame (a), top inset] and numerical [III, frame (d)] results are in qualitative agreement. The cross-correlation is time-asymmetric, and for positive delays it shows two clear periods of semiclassical oscillation with the same phase as in the waiting-time distribution $W_F$. The exact numerical results display the intermediate timescale -- being absent from the analytical expression -- which can also be discerned in the histograms. We conclude this panel sequence by reverting to direct photodetection. In Fig.~\ref{fig:fig8} IV (a), the intense revival of the quantum beat between emissions is superimposed on the semiclassical ringing; there is no sign of relaxation close to the steady-state average. The extracted waiting-time distribution for the forwards scattered light is more faithful to the ME and quantum regression formula predictions, yielding the same average time between emissions (compare also with the waiting-time distributions along the two channels numerically extracted in Ref.~\cite{Tian1992} for the driven ``vacuum'' Rabi resonance). 
 
Finally, let us look more closely at the fluctuations of the cavity field in response to single photon emission events under the two-photon resonance operation, starting with a balanced wave-particle correlator ($r=0.5$). In Fig.~\ref{fig:fig9}(a) we set $\theta=\pi/4$ and observe the variation of $\langle A_{\theta=\pi/4} (t) \rangle_{\rm REC}$ in a fragment of a single realization where seven quantum jumps are recorded. The initial state is a two-photon state. The Wigner function on the left portrays the cavity-field distribution at an instant when the conditional photon number attains a maximum after the first five decoherence times. It evinces a quantum superposition mainly between the two-photon Fock state and the steady-state profile of bimodality (see also the discussion on variants of photon-added states in Refs.~\cite{Zavatta2004, Abah2020}). The latter is also present in the asymmetry of the profile that resembles a single-photon phase-space distribution following a spontaneous emission (inset B). A sideways emitted photon initiates the quantum beat, as it also does in the conditioned field quadrature (see, {\it e.g.}, the field evolution following the emission at $\kappa t \approx 13.1$). Now there is no relaxation to a steady state, in contrast to the trajectory of Figs.~\ref{fig:fig8} I, II (a), owing to the stronger BHD backaction. 

Next, in frames (b) and (c) of Fig.~\ref{fig:fig9}, we work with a heavily unbalanced correlator [$95/5$ as in (a)] and a drive amplitude at which bimodality has not yet developed. There is however quadrature amplitude squeezing as well as strong photon bunching [$g^{(2)}(0) \approx 12.8$]. Here we meet the anomalous phase first reported experimentally in Ref.~\cite{Foster2000}. For both settings of $\theta$, along ($\pi/4$) and orthogonal ($3\pi/4$) to the direction of squeezing, we find that single jumps entail a clear time asymmetry; moreover, for $\theta=3\pi/4$, the conditioned field changes sign. For the latter, the symmetry shown in the wave-particle correlation function in the inset of Panel II in Fig.~\ref{fig:fig8} is recovered as a result of the BHD backaction, despite the fact that only 5\% of the incoming signal flux is subject to homodyne detection. The backaction is due to the quantum superposition of the local oscillator to the cavity signal field: a detection of a local-oscillator photon affects the evolution of the cavity field. The implication is that a conditional system state at a given time $t$ is correlated with the shot noise that has been generated in the recent past. Through the trigger clicks (particle aspect of light) a fraction of the shot noise is selected, the one being filtered through the system's dynamical response function~\cite{Reiner2001}.  
 
The two jumps at $\kappa t\approx 341.2$ are separated by a very small fraction of the cavity decay time [see Fig.~\ref{fig:figApp}(b)]; in fact, they are separated by half a beating period. The revival of the quantum beat timescale in the conditioned photon number acquires a further operational meaning in photon bunching conditions: it allows the experimenter to assess the time lapsed between emitted pairs in terms of the cycles completed within a couplet of the JC spectrum. At $\theta=3\pi/4$, the $\pi$ phase shift occurs when the conditioned photon number has a fluctuation up, in contrast to the classical intuition. At $\theta=\pi/4$, the field also has a maximum together with the conditioned probability. Having seen the disparity in the conditioned field (wave aspect) correlated with trigger counts (particle aspect), let us explore the reverse: particle counts correlated with the quadrature being measured. For that we appeal once again to the distribution of waiting times between emissions.

Figure~\ref{fig:fig9}(c) presents two waiting-time distributions (in the form of histograms) for two orthogonal quadratures particular to the redistribution of quantum fluctuations in the presence of strong photon antibunching. Waiting times between successive emissions are redistributed too for the two unravelings, which is more notable for short delays. The redistribution of waiting times among the selected quadratures testifies to the ability of the wave-particle correlator to reveal the contextual duality of light, despite that the inctracavity steady-state excitation is as small as $\langle a^{\dagger}a \rangle_{\rm ss}\approx 0.03$. As we approach the direction of anti-squeezing, successive photon emissions tend to be spaced further apart by a fraction of the cavity decoherence time with respect to the squeezing direction. Although the quantum beat is not properly resolved on the scale of the figure, both distributions evince a modulation which wanes off at time delays past the average photon lifetime.  

\section{Concluding remarks}

In summary, the conditional evolution of the photon number and the cavity field in the regime of photon blockade evinces a pronounced asymmetry of quantum fluctuations, revealed by the very function of the wave-particle correlator designed here to function with equal spontaneous emission and photon decay rates. One manifestation of such an asymmetry concerns the ratio between forwards and sideways emissions which, apart from being dependent on the selected multiphoton resonance determining the scattered field to be measured -- involving a particular subset of the JC excitation ladder-- also varies with quadrature of the electromagnetic field selected by the local oscillator. The asymmetry also concerns the evolution before and after a ``start" click as captured by the wave-particle (or amplitude-intensity) correlation function of the light emanating from a multiphoton resonance. The asymmetry is highly contextual, {\it i.e.}, it depends on the particular unraveling decided upon the choice of the local-oscillator phase, and may as well be tied in with the occurrence of a nonclassical phase shift for a manifestly negative spectrum of squeezing. There is a second aspect the asymmetry takes on, however, perhaps more subtle than the first: when operating the wave-particle correlator with a very small fraction of the flux directed towards the backaction of homodyne detection, the extracted statistical properties of the collected photons do not converge to those obtained for direct photodetection. Evidently, a larger number of conditioned homodyne-current samples $N_s$ (see App.~\ref{sec:thewp}) is required to asses this ``discontinuity'', which we leave to a future study. Nevertheless, the preliminary results reported here show that conditional probabilities as well as waiting times between emissions are highly sensitive to probing the dual nature of the light escaping the cavity in a multi-photon resonance operation, a field which, as we have seen, may be both bunched or antibunched depending on the operation settings.  

Detailed balance, operationally defined from measurements made through photodetection on outgoing fields, ostensibly breaks down as we traverse the series of the JC multiphoton resonances including the ``vacuum'' Rabi resonance. The latter is a special case which can be mapped to ordinary resonance fluorescence; even there, however, unambiguous deviations from a two-state scattering process are noted. The conditional cavity-field distributions, which can be operationally determined via mode-matched (homo/hetero)dyne detection, reveal a distinct quantum interference between single-photon states originating from the JC spectrum and the steady-state profile of quantum fluctuations fundamentally influenced by the inputs and outputs. In direct photodetection, the interference dynamically traces the cascade of emissions underlying an $n$-photon resonance, and is more akin to the steady-state profile once the density matrix has been averaged over a bundle of $n$ emissions. Is it degraded by the presence of the shot noise when operating the wave-particle correlator. 

Quantum interference is also recorded in the form of the quantum beat appearing in the field amplitude conditioned on a particular ``start" click. Once again, there is a disparity between the two types of emission, here in terms of their ability to initiate the quantum beat, carrying over from the ''vacuum'' Rabi resonance. For $n \geq 3$, the interference of quantum beats with different frequencies, itself present in the intensity correlation function of the forwards-scattered light~\cite{Mavrogordatos2022}, emerges as a dynamical feature conditioned on the number of photons emitted from the particular cascade process that substantiates a multi-photon resonance along the JC excitation ladder. Operationally determined Wigner distributions of the cavity field~\cite{CarmichaelKochan1994, Lutterbach1997, Nogues2000, CarmichaelQO2} are well suited to reveal both instantaneous and averaged dressed-state interferences as the cascaded decay pans out.  

I {\it acknowledge} edifying discussions with P. Rice. The work was supported by the Severo Ochoa Center of Excellence as well as by the Swedish Research Council (VR).

\appendix
\onecolumngrid
\setcounter{figure}{0} 
\renewcommand\thefigure{\thesection.\arabic{figure}} 
\section{Four-level model and derivation of the particle cross-correlation for the two-photon resonance; comparison with the ``vacuum'' Rabi resonance}
\label{sec:4levelcross}

In the interaction picture, the reduced density matrix $\rho$ of the open system evolves according to the standard Lindblad ME defining the two monitoring channels:
\begin{equation}\label{eq:ME}
 \frac{d\rho}{dt}\equiv \mathcal{L}\rho=-i[H_{\rm JC}/\hbar,\rho]+\kappa (2 a \rho a^{\dagger} -a^{\dagger}a \rho - \rho a^{\dagger}a)+\tfrac{\gamma}{2}(2\sigma_{-}\rho \sigma_{+} - \sigma_{+}\sigma_{-}\rho - \rho \sigma_{+}\sigma_{-}).
\end{equation}
Multiphoton resonances are excited between JC dressed states in the manifold comprising the ground state $|\xi_0\rangle=|0, -\rangle$ and the excited couplet states $|\xi_{n}\rangle=\tfrac{1}{\sqrt{2}}(|n, -\rangle-|n-1,+\rangle)$ and $|\xi_{n+1}\rangle=\tfrac{1}{\sqrt{2}}(|n, -\rangle+|n-1,+\rangle)$ for $n\geq 1$, where $|n,\pm\rangle \equiv |n\rangle |\pm \rangle$, with $|\pm\rangle$ the upper and lower states of the two-state atom and $\ket{n}$ the Fock states of the cavity field. For the minimal four-level model introduced in~\cite{Shamailov2010, Lledo2021}, the two-photon excitation involves the ground state $|\xi_0\rangle$ and the lower state of the second couplet $|\xi_3 \rangle$, while the cascaded decay is mediated by the first excited couplet states $|\xi_{1} \rangle, |\xi_{2}\rangle$. The mediation is observed through a coherent quantum beat~\cite{Shamailov2010}.  

To derive analytical results for steady-state correlation functions (two-time averages) we employ the minimal four-state model and the quantum regression formula~\cite{CarmichaelQO1} under the secular approximation~\cite{Cohen_Tannoudji_1977} operating in the strong-coupling limit of non-perturbative QED, $g \gg \kappa, \gamma/2$~\cite{CarmichaelQO2}. Adiabatically eliminating the intermediate states $|\xi_1 \rangle, |\xi_2\rangle$ but not discarding them from the dynamics, leads to the following effective ME [see Eq. (18) of Ref.~\cite{Shamailov2010}]
\begin{equation}\label{eq:ME2}
\begin{aligned}
  \frac{d\rho}{dt}=\tilde{\mathcal{L}}\rho \equiv& -(i/\hbar)[\tilde{H}_{\rm eff},\rho]+\Gamma_{32} \mathcal{D}[|\xi_2\rangle \langle \xi_3|](\rho)\\
  &+ \Gamma_{31} \mathcal{D}[|\xi_1\rangle \langle \xi_3|](\rho)
  + \Gamma \mathcal{D}[|\xi_0\rangle \langle \xi_1|](\rho) + \Gamma \mathcal{D}[|\xi_0\rangle \langle \xi_2|](\rho),  
  \end{aligned}
\end{equation}
with an effective Hamiltonian modeling the driving of the two-photon transition,
\begin{equation}
 \tilde{H}_{\rm eff}\equiv\sum_{k=0}^{3} \tilde{E}_{k} |\xi_k\rangle \langle \xi_k| + \hbar \Omega (e^{2i\omega_d t} |\xi_0\rangle \langle \xi_3| + e^{-2i\omega_d t} |\xi_3\rangle \langle \xi_0|).
\end{equation}
The intermediate states should not be disregarded since they take part in the cascaded process, with observational consequences. The JC energy levels with perturbative energy shifts (to second order in $\varepsilon_d/g$) are given below:
\begin{subequations}\label{eq:shifts}
\begin{align}
 &\tilde{E}_0=E_0 + \hbar\delta_0(\varepsilon_d) = \hbar \sqrt{2} \varepsilon_d^2/g, \\
 & \tilde{E}_1=E_1 + \hbar\delta_1(\varepsilon_d) = \hbar \{\omega_0 -  g - [(20 + 19\sqrt{2})/7]\varepsilon_d^2/g\}, \\
 & \tilde{E}_2=E_2 + \hbar\delta_2(\varepsilon_d) = \hbar \{\omega_0 +  g +  [(20 - 19\sqrt{2})/7]\varepsilon_d^2/g\}, \\
 & \tilde{E}_3=E_3 + \hbar\delta_3(\varepsilon_d) =  \hbar (2\omega_0 - \sqrt{2}  g -  \sqrt{2}\, \varepsilon_d^2/g),
\end{align}
\end{subequations}
and the effective two-photon coupling strength is $\Omega=2\sqrt{2}\, \varepsilon_d^2/g$~\cite{Lledo2021}. In the ME~\eqref{eq:ME2}, we define $\mathcal{D}[X](\rho)\equiv X\rho X^{\dagger}-(1/2)\{X^{\dagger}X, \rho\}$, while the transition rates between the four levels -- in the special case of impedance matching $\gamma=2\kappa$ -- are
\begin{subequations}
\begin{align}
 &\Gamma_{31}\equiv\tfrac{\gamma}{4}+(\sqrt{2}+1)^2 \tfrac{\kappa}{2}=\tfrac{\gamma}{4}[1+(\sqrt{2}+1)^2], \label{eq:Gamma31} \\
 & \Gamma_{32}\equiv\tfrac{\gamma}{4}+(\sqrt{2}-1)^2 \tfrac{\kappa}{2}=\tfrac{\gamma}{4}[1+(\sqrt{2}-1)^2], \\
 & \Gamma\equiv\tfrac{\gamma}{2}+\kappa=\gamma.
\end{align}
\end{subequations}
Including the perturbative energy shifts determined above, the two-photon resonance must be excited with a drive frequency $\omega_d$ satisfying $2\omega_d=(\tilde{E}_3-\tilde{E}_1)/\hbar=2\omega_0-\sqrt{2}g + \delta_3(\varepsilon_d)-\delta_0(\varepsilon_d)$, whence $\Delta\omega_d\equiv \omega_d-\omega_0=-g/\sqrt{2} - \sqrt{2}\, \varepsilon_d^2/g$. The effective ME~\eqref{eq:ME2} governs the evolution of the matrix elements in the dressed-state basis. To evaluate the second-order correlation functions we discussed in Sec.~\ref{sec:particlecorsscorr} it needs to be solved twice~\cite{CarmichaelQO1, CarmichaelQO2}; first to obtain the steady-state density matrix $\rho_{\rm ss}$ and second to advance the argument of the evolution up to the time delay $\tau$ subject to a photon emission which provides the conditioning. 

The system state conditioned on the ``first'' photon emission in the forwards direction (channel A) after steady state is attained is
\begin{equation}
 \rho_{\rm cond, 1}=\frac{a \rho_{\rm ss} a^{\dagger}}{\langle a^{\dagger} a \rangle}_{\rm ss}=\tfrac{2}{5}|\xi_0 \rangle \langle \xi_0| + \tfrac{3}{5}|\psi_{\rm super}\rangle \langle \psi_{\rm super}|,
\end{equation}
where $|\psi_{\rm super}\rangle$ is the quantum-beat superposition state
\begin{equation}
 |\psi_{\rm super}\rangle=\sqrt{\tfrac{2}{3}} \left(\tfrac{\sqrt{2}+1}{2}|\xi_1 \rangle +  \tfrac{\sqrt{2}-1}{2}|\xi_2 \rangle\right).
\end{equation}
Likewise, a ``first'' sideways-scattered photon (channel B) prepares the JC system in the state
\begin{equation}
 \rho_{\rm cond, 2}=\frac{\sigma_{-} \rho_{\rm ss} \sigma_{+}}{\langle \sigma_{+} \sigma_{-} \rangle}_{\rm ss}=\tfrac{2}{3}|\xi_0 \rangle \langle \xi_0| + \tfrac{1}{3}|\psi_{\rm super}\rangle \langle \psi_{\rm super}|.
\end{equation}
Expressing the two-level atomic excitation and photon number operator in the dressed-state basis, we write for the correlation AB with $\tau\geq 0$:
\begin{equation}
\text{Corr}_{AB}(\tau)={\rm tr} [\sigma_{+}\sigma_{-} e^{\mathcal{L}\tau}\rho_{\rm cond,1}]=\tfrac{1}{2}[\rho_{11}(\tau) + \rho_{22}(\tau) + \rho_{\rm 33}(\tau)-2{\rm Re}\{\rho_{12}(\tau)\}]
\end{equation}
and likewise, for $\tau\leq 0$,
\begin{equation}
\text{Corr}_{AB}(\tau)={\rm tr} [a^{\dagger}a e^{\mathcal{L}|\tau|}\rho_{\rm cond,2}]=\tfrac{1}{2}[\rho_{11}(|\tau|) + \rho_{22}(|\tau|) + 3\rho_{\rm 33}(|\tau|)+2{\rm Re}\{\rho_{12}(|\tau|)\}],
\end{equation}
where the matrix element evolve according the the ME~\eqref{eq:ME2} and are subject to the initial conditions $\rho(\tau=0^{\pm})=\rho_{{\rm cond}, 1}, \rho_{{\rm cond}, 2}$, respectively. The equations of motion for the matrix elements are solved by
\begin{equation}
 \rho_{33}=C e^{-2\gamma\tau} + \tfrac{\Omega^2}{\gamma^2 + 4\Omega^2} - \Sigma(0) e^{-\gamma \tau}\gamma \tfrac{\Omega}{\gamma^2 + 4\Omega^2}[\sin(2\Omega \tau)-2\tfrac{\Omega}{\gamma} \cos(2\Omega \tau)]- \tfrac{\Omega\gamma}{\gamma^2 + 4\Omega^2}e^{-\gamma \tau}\sin(2\Omega \tau).
\end{equation}
and
\begin{equation}
\begin{aligned}
  \rho_{11} +  \rho_{22}=& -2C e^{-2\gamma \tau} + 2\tfrac{\Omega^2}{\gamma^2 + 4\Omega^2} + 2\Sigma(0) e^{-\gamma \tau} \gamma\tfrac{\Omega}{\gamma^2 + 4\Omega^2}\left[\tfrac{\gamma}{2\Omega} \cos(2\Omega \tau)+ \sin(2\Omega \tau)\right]\\
  &+\tfrac{\gamma^2}{\gamma^2 + 4\Omega^2}e^{-\gamma \tau}\cos(2\Omega \tau),
  \end{aligned}
\end{equation}
with
\begin{equation}
 C=-\tfrac{\Omega^2}{\gamma^2 + 4\Omega^2} [1+2\Sigma(0)]\quad   \text{and}\quad \Sigma(0) \equiv \rho_{33}-\rho_{00},
\end{equation}
common to both conditioning emissions, reflecting the fact that $\rho_{33}(0^+)=0$ for any type of emission. The quantity $\Sigma(0)$ defines the effective ``inversion'' of the two-photon transition. As for the quantum beat, we obtain
\begin{equation}
 2{\rm Re}\{\rho_{12}\}=D e^{-\gamma \tau} \cos(\nu \tau).
\end{equation}
For a photon emission in transmission we get $\Sigma_1(0)=-\frac{2}{5}$ while for a sideways-scattered photon, $\Sigma_2 (0)=-\frac{2}{3}$, while the coefficients for the quantum beat read $D_1=\frac{1}{5}$ and $D_2=\frac{1}{3}$, respectively.

Putting the pieces together, and normalizing by the steady-state two-level atom excitation $\langle \sigma_{+}\sigma_{-} \rangle_{\rm ss}=\frac{3}{2}p_3$ and the steady-state cavity photon number $\langle a^{\dagger}a \rangle_{\rm ss}=\frac{5}{2}p_3$ [where $p_3=\Omega^2/(4\Omega^2+\gamma^2)$ is the steady-state occupation probability of $|\xi_3 \rangle$], we obtain the final expressions:
\begin{equation}\label{eq:crosscorrA1}
\begin{aligned}
 g^{(2)}_{AB}(\tau)\equiv \frac{\text{Corr}_{AB}(\tau)}{\langle \sigma_{+}\sigma_{-} \rangle_{\rm ss}}=& 1 + \tfrac{1}{15}e^{-2\gamma \tau} -\tfrac{7}{15}\tfrac{\gamma}{\Omega}e^{-\gamma \tau}\sin(2\Omega\tau) + \tfrac{1}{15}\left(3\tfrac{\gamma^2}{\Omega^2} -4\right)e^{-\gamma \tau}\cos(2\Omega\tau)\\
 &-\tfrac{1}{15}\tfrac{\gamma^2+4\Omega^2}{\Omega^2}e^{-\gamma\tau}\cos(\nu\tau), \quad \quad \text{for}\quad \tau\geq 0
 \end{aligned}
\end{equation}
and
\begin{equation}\label{eq:crosscorrA2}
\begin{aligned}
 g^{(2)}_{AB}(\tau)\equiv \frac{\text{Corr}_{AB}(\tau)}{\langle a^{\dagger}a \rangle_{\rm ss}}=& 1 + \tfrac{1}{15}e^{-2\gamma |\tau|} -\tfrac{7}{15}\tfrac{\gamma}{\Omega}e^{-\gamma |\tau|}\sin(2\Omega|\tau|) + \tfrac{1}{15}\left(\tfrac{\gamma^2}{\Omega^2} -12\right)e^{-\gamma |\tau|}\cos(2\Omega\tau)\\
 &+\tfrac{1}{15}\tfrac{\gamma^2+4\Omega^2}{\Omega^2}e^{-\gamma|\tau|}\cos(\nu\tau), \quad \quad \text{for}\quad \tau\leq 0.
 \end{aligned}
\end{equation}
We note the continuity at zero delay, $g^{(2)}_{AB}(0^+)=g^{(2)}_{AB}(0^-)$.

Two timescales coexist in the correlations of Eqs.~\eqref{eq:crosscorrA1},~\eqref{eq:crosscorrA2}. The fastest corresponds to the quantum beat, with frequency equal to $\nu \equiv (\tilde{E}_2-\tilde{E}_1)/\hbar=2g + \delta_2-\delta_1$. The slowest corresponds to the semiclassical ringing with frequency $2\Omega=4\sqrt{2}\varepsilon_d^2/g$~\cite{Lledo2021} and is associated with driving a saturable (effective) two-level transition, similar to ordinary resonance fluorescence. An intermediate timescale, which is missed in the secular approximation and therefore in the analytical treatment, is present in the numerically-determined cross-correlation plotted in Fig.~\ref{fig:fig8}III(d). It is due to the asymmetry between the excitation paths $|\xi_3\rangle \to |\xi_1\rangle \to |\xi_0\rangle$ and $|\xi_3\rangle \to |\xi_2\rangle \to |\xi_0\rangle$, since $\Gamma_{31}/\Gamma_{32}=5.8$ for $\gamma/(2\kappa)=0$~\cite{Shamailov2010}. In a frame rotating with the drive and for $\Delta\omega_d=-g/\sqrt{2}$, the unperturbed energies of the intermediate states are $E^{(0)}_{1,2}=g(1\mp \sqrt{2})/\sqrt{2}$ whereas the outer states $|\xi_3 \rangle$ and $|\xi_0\rangle$ have zero energy. It follows that the excitation path $|\xi_3\rangle \to |\xi_1\rangle \to |\xi_0\rangle$ is associated with a periodic structure including $2\sqrt{2}/(\sqrt{2}-1)\approx 6.8$ quantum beat cycles while the path $|\xi_3\rangle \to |\xi_2\rangle \to |\xi_0\rangle$ is associated with a period of about $1.2$ times the beat cycle, manifested as an occasional deviation from the pattern dictated by the dominant path. 

When assessing the particle aspect of the forward- and side-scattered light, we also employ the waiting-time distribution defined by conditional exclusive probability densities. For forward scattering, this quantity is given by~\cite{Carmichael1989, Carmichael1993}
\begin{equation}
 w_{F}(\tau)=2\kappa \frac{{\rm tr}[a^{\dagger}a e^{\overline{\mathcal{L}} \tau}(a \rho_{\rm ss}a^{\dagger})]}{\braket{a^{\dagger}a}_{\rm ss}}\\
 =(2\kappa)\,{\rm tr}\{a^{\dagger}a e^{\overline{\mathcal{L}} \tau} [\rho_{\rm cond}]\},
\end{equation}
where $\overline{\mathcal{L}} \equiv \mathcal{L}-2\kappa a \cdot a^{\dagger}$ and $\rho_{\rm cond} \equiv (a \rho_{\rm ss} a^{\dagger})/\braket{a^{\dagger}a}_{\rm ss}$ is the (normalized) conditioned density matrix following the post-steady-state emission of the ``first'' photon. The intensity correlation function is instead given by~\cite{Carmichael1989, Carmichael1993} $g^{(2)}(\tau)={\rm tr}\{a^{\dagger}a e^{\mathcal{L} \tau} [\rho_{\rm cond}]\}/\braket{a^{\dagger}a}_{\rm ss}$, corresponding to an unconditional probability (the Liouvillian $\mathcal{L}$ replaces $\overline{\mathcal{L}}$).

All numerical results involving unconditional dynamics (ME~\eqref{eq:ME} and the quantum regression formula) were obtained in \textsc{Matlab}'s {\it Quantum Optics Toolbox}~\cite{Tan_1999} using an exponential series expansion of the density operator in a Hilbert space for the cavity mode truncated at the $14^{\rm th}$ photon level. An example of the developing temporal asymmetry in the intensity cross-correlation as the incoherent scattering involves higher rungs along the JC ladder is illustrated in Fig.~\ref{fig:figAppCrossCorr}.

\begin{figure*}
\centering
 \includegraphics[width=0.5\textwidth]{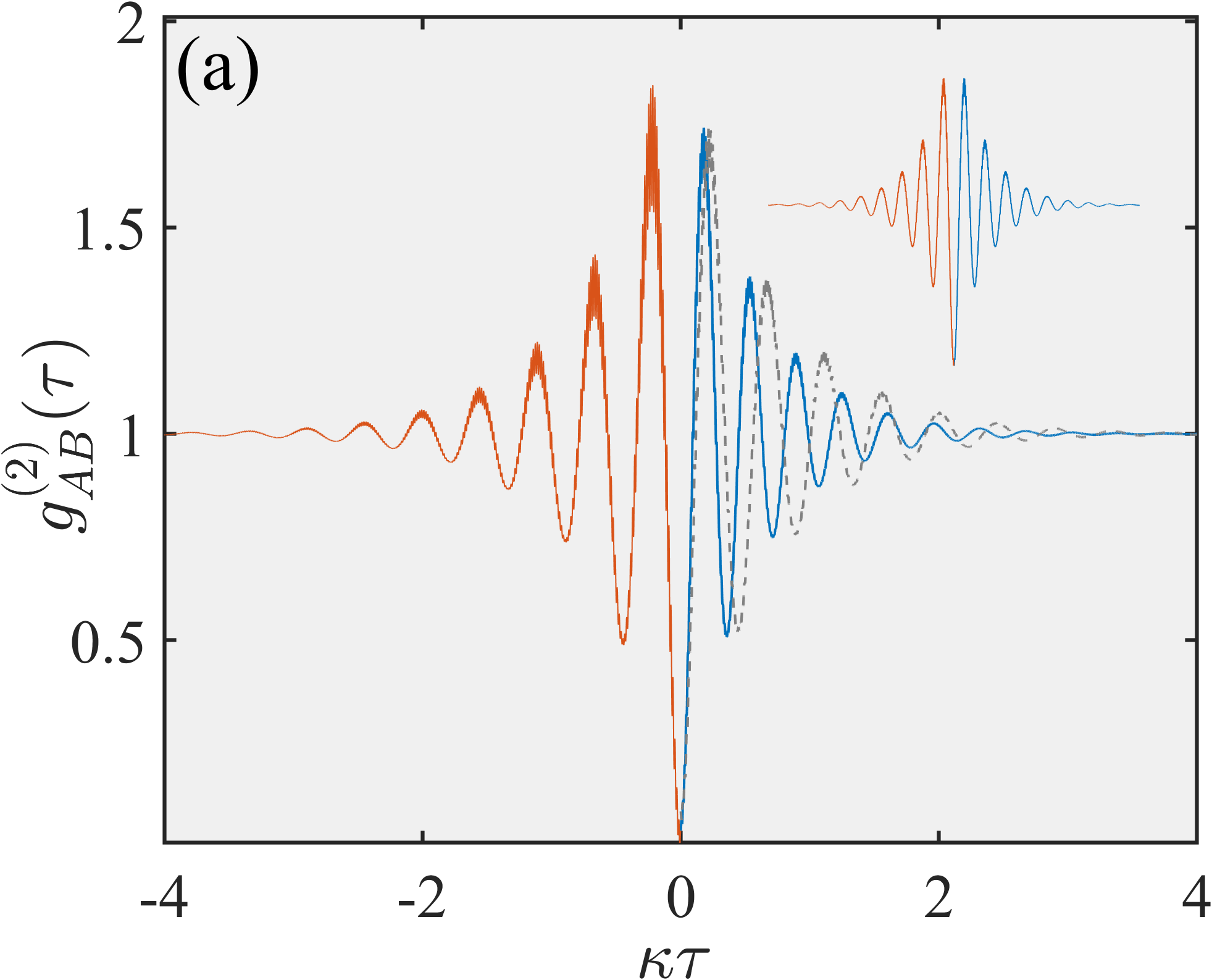}
  \includegraphics[width=0.475\textwidth]{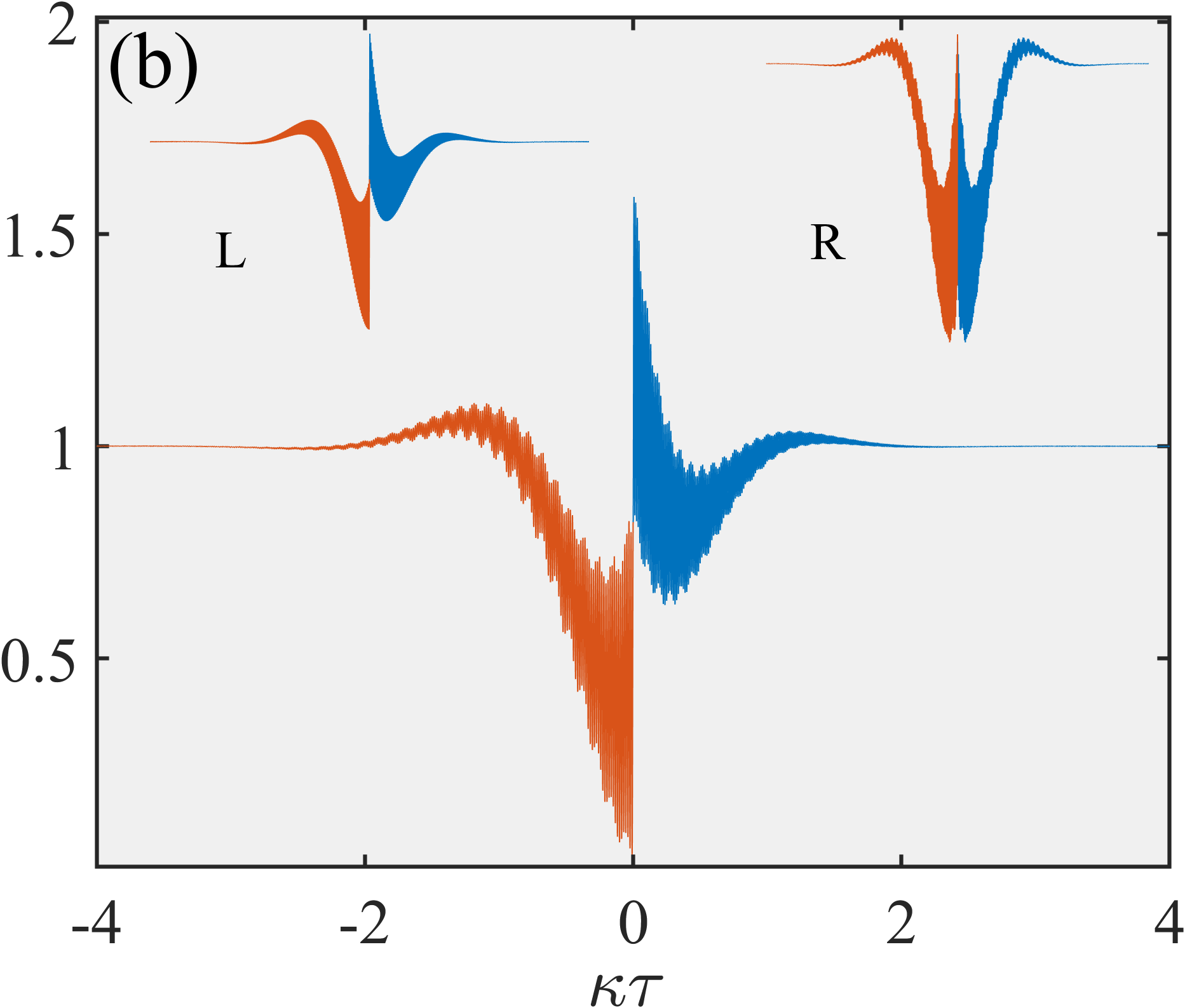}
  \caption{Intensity cross-correlation recorded for the two different channels in the course of scattering at the ``vacuum'' Rabi {\bf (a)} and two-photon {\bf (b)} resonance. The function $g_{AB}(\tau)$ is calculated from the numerical solution to the ME~\eqref{eq:ME} and the quantum regression formula, for $g/\kappa=200$, $\varepsilon_d/g=0.05$ and $|\Delta\omega_d/g|=1$ in {\bf (a)}; $|\Delta\omega_d/g|=1/\sqrt{2} + \sqrt{2}(\varepsilon_d/g)^2$ in {\bf (b)}. The dashed grey line in frame (a) overlays the intensity correlation function $g^((2))(\tau)$ for $\tau \geq 0$. The left (L) inset in frame (b) depicts the analytically-derived cross-correlation from Eq.~\eqref{eq:crosscorr}, while the right (R) inset depicts the numerically-extracted intensity (auto-)correlation $g^{2}(\tau)$ for the same parameters.}
  \label{fig:figAppCorr}
\end{figure*}

\begin{figure*}
\centering
 \includegraphics[width=\textwidth]{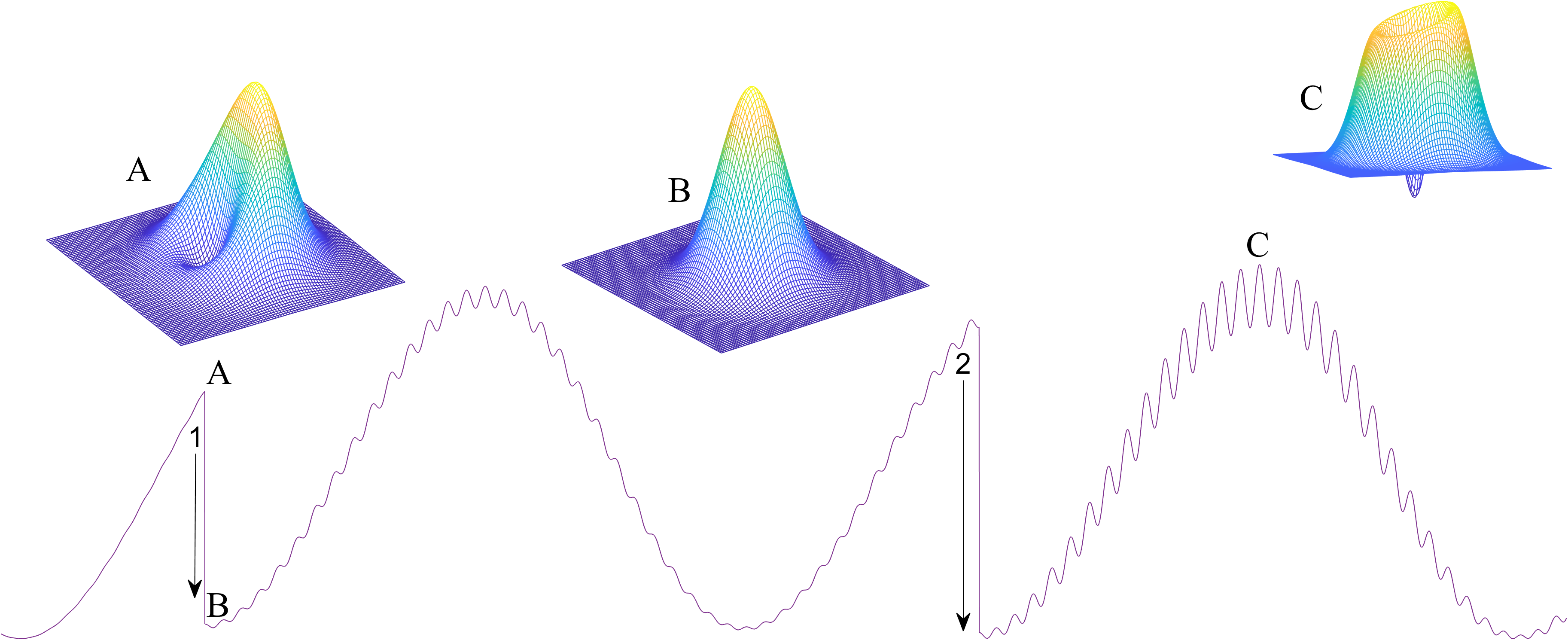}
  \caption{Fraction of a sample realization $\langle a^{\dagger}a (t) \rangle_{\rm REC}$ in the time interval $27.7 \leq \kappa t \leq 28.8 $ when driving at the ``vacuum'' Rabi resonance with the same parameters as in Fig.~\ref{fig:figAppCorr}(a). The three insets A, B, C depict conditioned Wigner distributions of the cavity field corresponding to the three marked instants of time: A, B resolve a cavity emission, while C is the position of the maximum in the second Rabi cycle on top of the quantum beat. Compare also with Fig. 1 of~\cite{Tian1992}.}
  \label{fig:figVR}
\end{figure*}

Figure~\ref{fig:figVR} depicts the conditional photon number during one cavity lifetime in the course of a sample trajectory for a saturated ``vacuum'' Rabi transition with $\langle a^{\dagger} a \rangle_{\rm ss}=1/4$~\cite{Tian1992}. Jump 1 corresponds to the emission of a cavity photon, while Jump 2 is a spontaneous emission event. The asymmetry in the conditioned Wigner function of inset A, just before the jump, is due to the finite off-diagonal elements $\langle 1| \rho_{\rm cav} | 0 \rangle$ and its conjugate, where $\rho_{\rm cav}$ is the cavity density matrix, and $|0\rangle, |1\rangle$ are the vacuum and single-photon states, respectively. The Wigner function of inset C resolves a 55/45 asymmetry in the participation of $\langle 0| \rho_{\rm cav} | 0 \rangle$ and  $\langle 1| \rho_{\rm cav} | 1 \rangle$ in the cavity density matrix; an imaginary part in  $\langle 1| \rho_{\rm cav} | 0 \rangle$ breaks the azimuthal symmetry. Spontaneous emissions initiate a higher-amplitude superposition of the quantum beat -- whose very appearance marks a departure from the two-state manifold -- as they result in a conditioned state with an increased proportion of $\langle 0| \rho_{\rm cav} | 0 \rangle$ compared to cavity emissions. This instance breaks the equivalence between atomic and cavity operators, which are treated on an equal footing in the analytical two-state approximation of the ``vacuum'' Rabi resonances (see  Sec. 13.3.3 of Ref.~\cite{CarmichaelQO2}). 

\setcounter{figure}{0} 
\section{The wave-particle correlator unraveling}
\label{sec:thewp}

We write the ME for the photoemissive source -- a driven JC oscillator operated in the photon blockade regime -- extended now to incorporate the local oscillator mode tuned to the frequency of the coherent drive $\omega_d$. The density matrix $\tilde{\rho}$ of the composite system in a frame rotating with $\omega_d$ reads~\cite{Reiner2001}:
\begin{equation}\label{eq:MEcorr}
\begin{aligned}
 \frac{d{\tilde{\rho}}}{dt}=&-i[H_{\rm JC}/\hbar,\tilde{\rho}]+\kappa (2 a \tilde{\rho} a^{\dagger} -a^{\dagger}a \tilde{\rho} - \tilde{\rho} a^{\dagger}a) + (\gamma/2)(2\sigma_{-}\tilde{\rho}\sigma_{+}-\sigma_{+}\sigma_{-}\tilde{\rho}-\tilde{\rho}\sigma_{+}\sigma_{-})\\
 &+\kappa_{\rm LO}(2c \tilde{\rho} c^{\dagger}-c^{\dagger}c\tilde{\rho}-\tilde{\rho}c^{\dagger}c) + \kappa_{\rm LO} \beta[c^{\dagger}-c,\tilde{\rho}],
 \end{aligned}
\end{equation}
where $c$ and $c^{\dagger}$ are the creating and annihilation operators for the local oscillator driven mode with excitation amplitude $\beta$. Under the {\it conditioned} balanced homodyne detection (BHD) scheme of Fig.~\ref{fig:fig1}, the total measured fields at the photodetectors 1 and 2 (in photon flux units) are written as  
\begin{equation}\label{eq:fielddet}
 \mathcal{E}_{{\rm BHD}_{1,2}}=\pm i \sqrt{\kappa_{\rm LO}}\,c + \sqrt{\kappa(1-r)}\,a,
\end{equation}
while the field measured at the avalanche photodiode (APD) is
\begin{equation}
 \mathcal{E}_{\rm count}=\sqrt{2\kappa r}\, a.
\end{equation}
Assuming that the local oscillator mode is unaffected by the signal mode, the density matrix factorizes into a part corresponding to the driven JC oscillator, operating here in the regime of photon blockade, and a term related to the local oscillator, $\tilde{\rho}=\tilde{\rho}_s |\beta\rangle \langle \beta|$.
We define the local oscillator flux as $f\equiv \kappa_{\rm LO}|\beta|^2$ and we construct the following two collapse super-operators acting on $\tilde{\rho}_s$:
\begin{subequations}\label{eq:supops}
 \begin{align}
\mathcal{S}_{{\rm BHD}_{1,2}}\tilde{\rho}_s&=(\pm \sqrt{f} e^{i\theta} + \sqrt{2\kappa(1-r)}\,a)\,\tilde{\rho}_s\,(\pm \sqrt{f} e^{-i\theta} + \sqrt{2\kappa(1-r)}\,a^{\dagger})\label{eq:supopsBHD},\\
\mathcal{S}_{\rm count}\tilde{\rho}_s&=2\kappa r\, a \tilde{\rho}_s a^{\dagger} \label{eq:supopscounts}.
 \end{align}
\end{subequations}
Subtracting the action of super-operators~\eqref{eq:supops}, defining the conditioned homodyne detection, from the Liouvillian of Eq.~\eqref{eq:MEcorr}, we arrive at the following expression for the super-operator which dictates the evolution of the un-normalized {\it conditioned} density operator $\tilde{\rho}_s (t)$:
\begin{equation}\label{eq:sup2det}
(\mathcal{L}-\mathcal{S}_{{\rm BHD}_{1,2}}-\mathcal{S}_{\rm count})\tilde{\rho}_s=-i[H_{\rm JC}/\hbar,\tilde{\rho}_s]-\kappa(a^{\dagger}a \tilde{\rho}_s + \tilde{\rho}_s a^{\dagger}a) - f\tilde{\rho}_s.
\end{equation}
Equation~\eqref{eq:sup2det} defines a non-Hermitian Hamiltonian which propagates the conditioned wavefunction $|\psi_{\rm REC}(t)\rangle$ of the system between measurements whose action is defined by the super-operators~\eqref{eq:supops}.

Two types of detections occur under the {\it conditioned} BHD measurement scheme. The first is modeled by Eq.~\eqref{eq:supopscounts} and corresponds to detection ``clicks'' at the APD. These collapses take place with probability $2\kappa r \langle \psi_{\rm REC}(t)|a^{\dagger}a|\psi_{\rm REC}(t)\rangle dt$. The second type of jumps has to do with the ``clicks'' registered via the BHD. In any realistic homodyne measurement the local oscillator photon flux $f$ is many orders of magnitude larger than the signal flux. The latter is $2\kappa (1-r) \langle \psi_{\rm REC}(t)|a^{\dagger}a|\psi_{\rm REC}(t)\rangle$. Consequently, under the action of~\eqref{eq:fielddet}, a photoelectric emission corresponds with high probability to an annihilation of a local oscillator photon. while there is only a small probability $~f/[2\kappa \langle \psi_{\rm REC}(t)|a^{\dagger}a|\psi_{\rm REC}(t)\rangle]$ that a photon originated from the cavity. We should note here that these two possibilities (co)-exist as a superposition and not as a classical choice, {\it either} one {\it or} the other. Despite the fact that the collapses are very small indeed, they occur very often on the characteristic timescale $(2\kappa)^{-1}$ for fluctuations in the signal field. The quantum mapping into a stochastic differential equation -- a Schr\"{o}dinger equation with a stochastic non-Hermitian Hamiltonian -- involves a coarse graining in time, an expansion of the non-unitary evolution in powers of $\sqrt{2\kappa(1-r)/f}$ and the consideration of a stochastic process for the number of the emitted photoelectrons which depends on a conditioned wavefunction that satisfies the very same Schr\"{o}dinger equation. Taking that path leads to the following expressions for the difference in photocurrents (normalized by the product $Ge |\varepsilon_{\rm lo}|$) between detectors 1 and 2 (see Fig.~\ref{fig:fig1}), and the wavefunction propagation between photodetections at the APD: 
\begin{equation}\label{eq:di}
di=-B (i\,dt - \sqrt{8\kappa(1-r)}\langle A_{\theta} \rangle_{\rm REC}\,dt - dW_t),
\end{equation}
\begin{equation}\label{eq:dpsi}
 d|\overline{\psi}_{\rm REC}\rangle=\bigg[\frac{1}{i\hbar}H^{\prime}\,dt + \sqrt{2\kappa(1-r)}\, a e^{-i\theta}(\sqrt{8\kappa(1-r)}\langle A_{\theta}(t) \rangle_{\rm REC}\,dt + dW_t)\bigg]|\overline{\psi}_{\rm REC}\rangle.
\end{equation}
Here $|\overline{\psi}_{\rm REC}\rangle$ is the un-normalized wavefunction and $H^{\prime}$ is the non-Hermitian Hamiltonian
\begin{equation}\label{eq:nonH}
 H^{\prime}\equiv H_{JC}-i\hbar \kappa a^{\dagger}a-i\hbar(\gamma/2)\sigma_{+}\sigma_{-}=-\hbar \Delta\omega_d (a^{\dagger}a + \sigma_+ \sigma_-)+ \hbar g (a^{\dagger}\sigma_- + a \sigma_+)+\hbar\varepsilon_d(a+a^{\dagger})-i\hbar \kappa a^{\dagger}a-i\hbar(\gamma/2)\sigma_{+}\sigma_{-}.
 \end{equation}
In Eqs.~\eqref{eq:di} and~\eqref{eq:dpsi}, $\langle A_{\theta}(t) \rangle_{\rm REC}$ is the conditioned average of the intracavity field $ \langle A_{\theta}(t) \rangle_{\rm REC} \equiv \tfrac{1}{2} \langle \psi_{\rm REC}(t)|e^{i\theta}a^{\dagger} + a e^{-i\theta}|\psi_{\rm REC}(t) \rangle$,
calculated with the normalized conditioned wavefunction $|\psi_{\rm REC} (t)\rangle$, $B$ is the detection bandwidth [{\it e.g.}, with filtering of an $RC$ circuit where $B=1/(RC)$] and $dW_t$ is the Wiener noise increment, the same in both equations~\eqref{eq:di} and~\eqref{eq:dpsi}. 
\begin{figure*}
\centering
 \includegraphics[width=\textwidth]{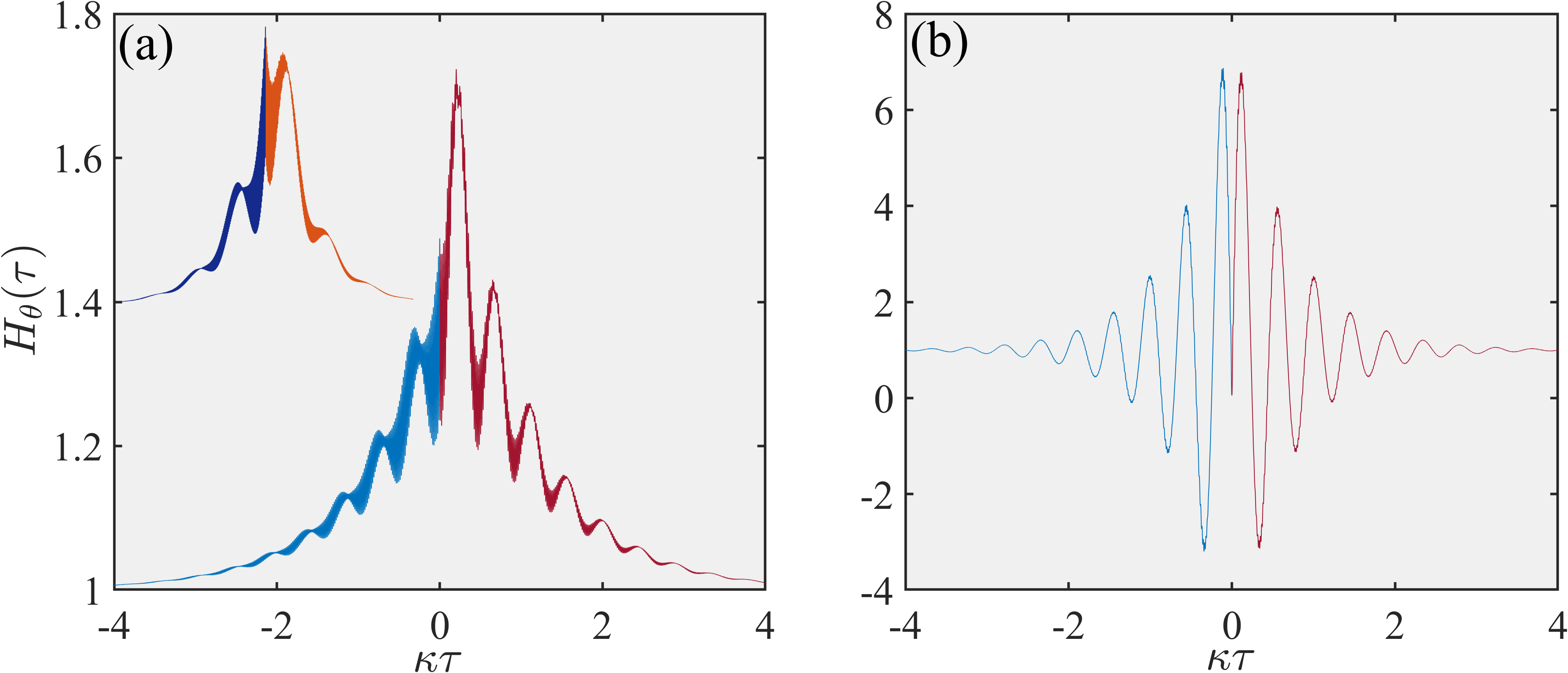}
  \caption{Normalized wave-particle correlation $H_{\theta}(\tau)$ following scattering at the ``vacuum'' Rabi resonance for $\theta=0$ in {\bf (a)} and $\theta=\pi/2$ in {\bf (b)}. The correlation is calculated from the numerical solution to the ME~\eqref{eq:ME} and the quantum regression formula, for $g/\kappa=200$, $\varepsilon_d/g=0.05$ [$\varepsilon_d/g=0.02$ for the upper left inset in (a)] and $|\Delta\omega_d/g|=1$.}
  \label{fig:figAppCrossCorr} 
\end{figure*}

The record obtained by the application of quantum trajectory theory comprises the continuous homodyne current $i(t)$ and a set of start times $\{t_j\}$. We sample an ongoing realization of $i(t)$ over many start times $N_s\gg 1$ to calculate the quantity~\cite{GiantViolations2000, CarmichaelFosterChapter}
\begin{equation}\label{eq:HNs}
 H_{\theta}(\tau)=\frac{1}{N_s}\sum_{j=1}^{N_s} i(t_j+\tau).
\end{equation}
Taking the limit $N_s\to \infty$ eliminates the residual local oscillator shot noise due to the sampling procedure and recovers Eq.~\eqref{eq:Hdef} normalized by the steady-state photon number. The latter is equal to $\langle A_{\theta}(\tau) \rangle_{c}$, where the subscript $c$ denotes a conditioning of the state on a photon emission at time $t$~\footnote{Throughout this report, for simplicity, we adopt the same symbol, $H_{\theta}(\tau)$, for the wave-particle correlation of different normalizations; for instance, the correlations of Fig.~\ref{fig:fig3} are normalized to unity, whereas the limit of~\eqref{eq:HNs} for $N_s \to \infty$ and a high detection bandwidth is normalized to a value proportional to the steady-state field amplitude.}. We also note that the correlation of the residual noise scales with the ratio $B/N_s$, while a large detection bandwidth is required ($B/\kappa \gg 1$) to resolve the source fluctuations in time~\cite{GiantViolations2000}. Realizations of $i(t)$, $\{t_j\}$ and $|\psi_{\rm REC}(t)\rangle$ are governed by the set of stochastic differential equations~\eqref{eq:di} and~\eqref{eq:dpsi}, solved here by means of an explicit order 2.0 weak scheme proposed by Kloeden and Platen~\cite{KloedenPlaten}. Spontaneous emission is also present and included in the standard way through additional quantum jumps.  

The intensity cross-correlation of side-scattered and forwards-scattered light offers a partial view on the symmetry of quantum fluctuations. Figure~\ref{fig:figAppCrossCorr} provides further evidence on the breakdown of detailed balance for the ``vacuum'' Rabi resonance. The (scaled) semiclassical oscillation frequency is $\sqrt{2} (\varepsilon_d/\kappa)$, the Rabi frequency for the "two-level system" identified in~\cite{Tian1992}. Nevertheless, the temporal asymmetry of the field-intensity correlation along the direction $\theta=0$ (X-quadrature) marks a clear departure from a two-state scattering process, accompanied by a more intense quantum beat, due to the participation of states up to $|\xi_3\rangle$. The profile becomes more symmetric for an increasing timespan about zero as the drive amplitude is lowered; symmetry is also restored when the local oscillator selects the orthogonal Y-quadrature. 

\begin{figure*}
\centering
 \includegraphics[width=0.475\textwidth]{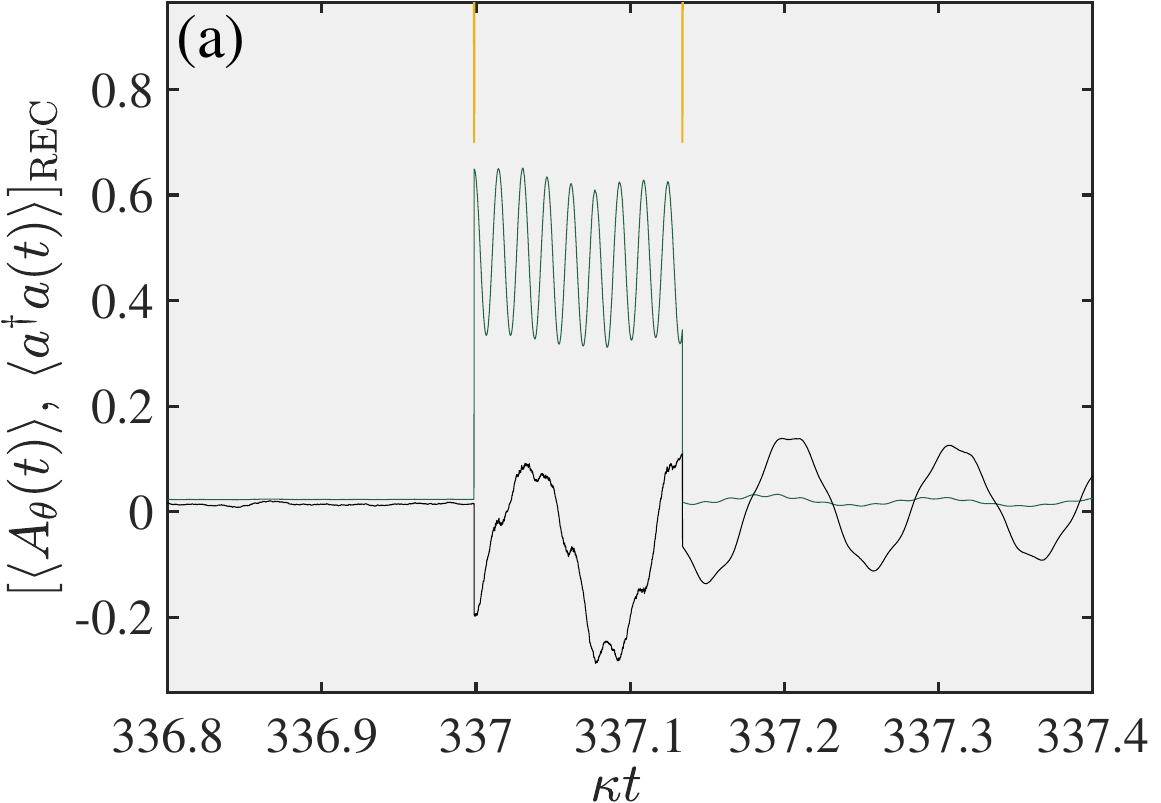}
  \includegraphics[width=0.49\textwidth]{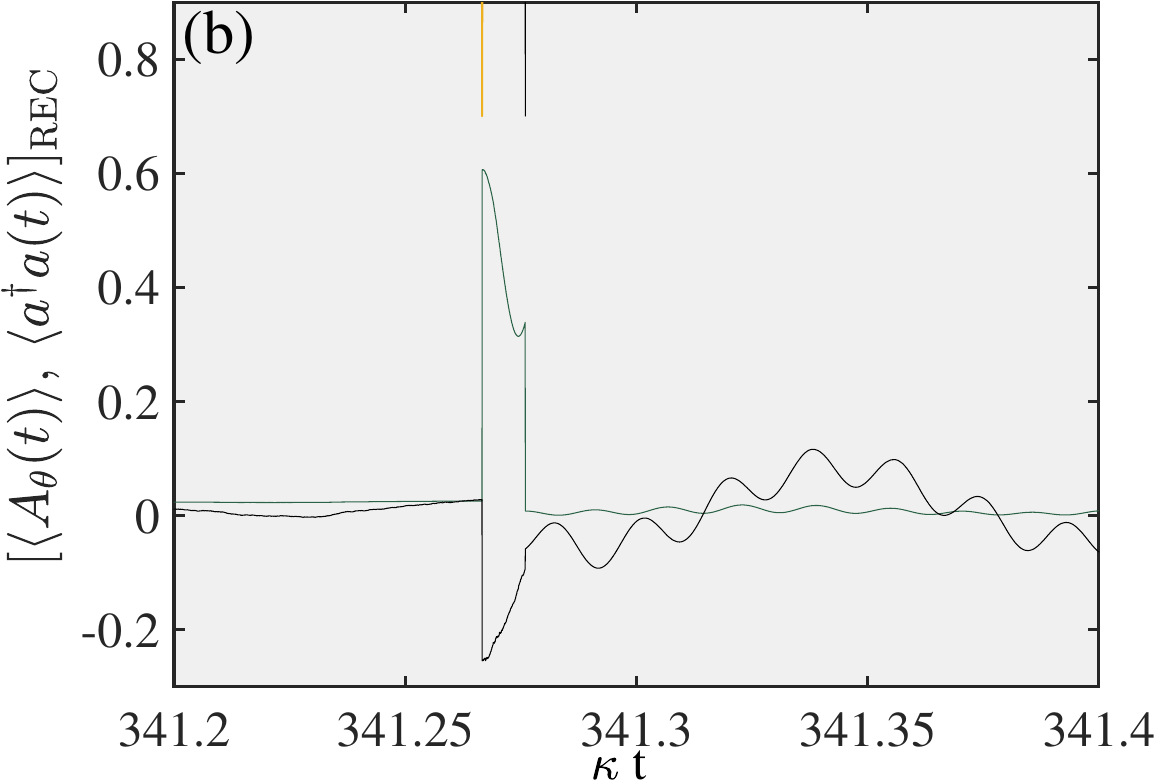}
  \caption{Focus on the two instances of anomalous phase switching marked by arrows in the trajectory of Fig.~\ref{fig:fig9}(b) generated for $\theta=3\pi/4$. Orange (black) strokes indicate cavity (spontaneous) emissions.}
  \label{fig:figApp}
\end{figure*}
Coming now to the excitation of the two-photon resonance, a fragment of a conditional realization depicting the anomalous phase switching of the field amplitude is depicted in Fig.~\ref{fig:figApp}. The field changes sign both when a photon escapes though the cavity mode while at the same time the conditional photon emission probability jumps upwards and the quantum beat is initiated. 

\section{Heterodyne detection of the cavity field}
\label{sec:hetdet}

In the conditional homodyne detection scheme of the wave-particle correlator, the frequency $\omega$ of the local oscillator matches the drive frequency $\omega_d$, which is at the center of the source-field spectrum. Heterodyne detection employs a local oscillator that is far detuned from the drive frequency. This brings in the detuning factors $e^{\pm \Delta\omega t}$, with $\Delta\omega \gg \kappa$, in the interference terms between the local oscillator and the source field. This means that effectively the local oscillator phase $\theta$ is replaced by $-\Delta\omega\, t$. In place of the stochastic Schr\"{o}dinger equation~\eqref{eq:dpsi} (taken with $r=0$), we now obtain:
\begin{equation}\label{eq:Het1}
  d|\overline{\psi}_{\rm REC}\rangle=\bigg[\frac{1}{i\hbar} H^{\prime}\,dt + \sqrt{2\kappa}\, a e^{i\Delta \omega t} dq\bigg]|\overline{\psi}_{\rm REC}\rangle,
\end{equation}
with
\begin{equation}\label{eq:Het2}
 dq=\sqrt{2\kappa}(\langle e^{-i\Delta\omega t}a^{\dagger} + e^{i\Delta\omega t}a \rangle_{\rm REC}\,dt + dW_t)
\end{equation}
the elemental charge deposited on the detector circuit (normalized by the product $Ge |\varepsilon_{\rm lo}|$). The non-Hermitian Hamiltonian $H^{\prime}$ is defined in Eq.~\eqref{eq:nonH}. Since the frequency mismatch $\Delta \omega$ is much larger than the bandwidth of the source-field fluctuations, we can introduce the slowly-varying differential charge: 
\begin{equation}
 d\tilde{q} \equiv e^{i\Delta\omega t} dq
\end{equation}
and neglect the rapidly oscillating term $\sqrt{2\kappa}\,
e^{2i\Delta\omega\, t}\langle a \rangle_{\rm REC}$ on the right-hand side of Eq.~\eqref{eq:Het1}. After making the substitution
\begin{equation}
 e^{i\Delta\omega\, t} dW_t \to dZ_t, \quad \text{with} \, dZ_t=(dW_{x;t} + i dW_{y;t})\sqrt{2} \,\, \text{and covariances}\,\,\overline{dZ_t dZ_t}=\overline{dZ_t^{*} dZ_t^{*}}=0, \quad \overline{dZ_t dZ_t^{*}}=dt, 
\end{equation}
we arrive at the following stochastic Schr\"{o}dinger equation within quantum trajectory theory for the open driven JC model with heterodyne current records:
\begin{equation}\label{eq:Het3}
  d|\overline{\psi}_{\rm REC}\rangle=\bigg[\frac{1}{i\hbar} H^{\prime}\,dt + \sqrt{2\kappa} a\, d\tilde{q}\,\bigg]|\overline{\psi}_{\rm REC}\rangle,
\end{equation}
where
\begin{equation}\label{eq:Het4}
 d\tilde{q}=\sqrt{2\kappa}(\langle a^{\dagger} \rangle_{\rm REC}\,dt + dZ_t).
\end{equation}
The stochastic process $d\tilde{q}$ has bandwidth $1/dt$. The complex heterodyne current is once again constructed with a realistic detection bandwidth $B$ (equal to $1/(RC)$ for an RC circuit) as
\begin{equation}
 d\tilde{i}=-B(\tilde{i}\,dt-d\tilde{q}).
\end{equation}
The raw heterodyne current is real and takes the form of a modulated RF carrier oscillation with a carrier frequency equal to the mismatch $\Delta \omega$. The complex current $\tilde{i}(t)$ accounts for the slowly-varying amplitude and phase of that modulation~\cite{CarmichaelQO2}. 

\twocolumngrid

\bibliography{Bibliography_DB}

\end{document}